\documentclass[english,aps,manuscript,aip,jcp,english,dvipsnames,preprint]{revtex4-1}
\usepackage[T1]{fontenc}
\usepackage[latin9]{inputenc}
\usepackage{amsbsy}
\usepackage{amstext}
\usepackage{amssymb}
\usepackage{graphicx}
\usepackage{esint}
\usepackage{subscript}

\makeatletter

\providecommand{\tabularnewline}{\\}

 \@ifundefined{textcolor}{}
 {%
   \definecolor{BLACK}{gray}{0}
   \definecolor{WHITE}{gray}{1}
   \definecolor{RED}{rgb}{1,0,0}
   \definecolor{GREEN}{rgb}{0,1,0}
   \definecolor{BLUE}{rgb}{0,0,1}
   \definecolor{CYAN}{cmyk}{1,0,0,0}
   \definecolor{MAGENTA}{cmyk}{0,1,0,0}
   \definecolor{YELLOW}{cmyk}{0,0,1,0}
 }

\def\part#1{\left(#1\right)}

\usepackage{amsthm}
\usepackage{graphics}
\usepackage{times}
\usepackage{bm}
\usepackage{hyperref}
\usepackage{color}
\usepackage{braket}

\setcitestyle{super}

\RequirePackage[dvipsnames]{xcolor} 

\hypersetup{
colorlinks=true,
urlcolor=Black, 
citecolor=Black,
linkcolor=Black 
}

\makeatother

\usepackage{babel}
\begin{document}

\title{``Divide-and-Conquer'' Semiclassical Molecular Dynamics: An Application
to Water Clusters}

\author{Giovanni \surname{Di Liberto}}

\affiliation{Dipartimento di Chimica, Università degli Studi di Milano, via C.
Golgi 19, 20133 Milano, Italy}

\author{Riccardo \surname{Conte}}

\affiliation{Dipartimento di Chimica, Università degli Studi di Milano, via C.
Golgi 19, 20133 Milano, Italy}

\author{Michele \surname{Ceotto}}
\email{michele.ceotto@unimi.it}

\affiliation{Dipartimento di Chimica, Università degli Studi di Milano, via C.
Golgi 19, 20133 Milano, Italy}
\begin{abstract}
We present an investigation of vibrational features in water clusters
performed by means of our recently established divide-and-conquer
semiclassical approach (M. Ceotto, G. Di Liberto, R. Conte \textit{Phys.
Rev. Lett. }\textbf{119}\textit{, }010401 (2017)). This technique
allows to simulate quantum vibrational spectra of high dimensional
systems starting from full-dimensional classical trajectories and
projection of the semiclassical propagator onto a set of lower dimensional
subspaces. The potential energy surface employed is a many-body representation
up to three-body terms, in which monomers and two-body interactions
are described by the high level WHBB water potential, while, for three-body
interactions, calculations adopt a fast permutationally-invariant
ab initio surface at the same level of theory of the WHBB 3-body potential.
Applications range from the water dimer up to the water decamer, a
system made of 84 vibrational degrees of freedom. Results are generally
in agreement with previous variational estimates in the literature.
This is particularly true for the bending and the high-frequency stretching
motions, while estimates of modes strongly influenced by hydrogen
bonding are red shifted, in a few instances even substantially, as
a consequence of the dynamical and global picture provided by the
semiclassical approach.
\end{abstract}
\maketitle

\section{Introduction\label{sec:Introduction}}

The water molecule has often attracted the attention of the scientific
community due to the fundamental role it plays for life on our planet.\citep{Wu_Voth_protonchannel_2003,smondyrev_Voth_phospholipid_2002,petersen_Voth_pocket_2005,mohammed_Nibbering_waterbrideges_2005,riccardi_Cui_protonholes_2006,wolke_Jhonson_snapshotsprotontransfer_2016,marx_abinitio_200years_2006}
In chemical and physical processes involving water, hydrogen bonding
is crucial\citep{Feyereisen_Dixon_MP2hydrogenbonden_1996,Poole_Angel_Hydrogenbondthermod_1994}
allowing the formation of supra-molecular structures (made of several
water molecules) known as water clusters.\citep{liu1_Clary_Hydrogenbondinghexamer_1996,xantheas_Hydrogenbonding_Abintio_2000,liu_saykally_waterclusters_1996}
Water clusters are major players in atmospheric photocatalytic processes,\citep{pfeilsticker_Bosh_waterdimeratmosphere_2003,vohringer_Abel_OHradicals_2007}
and they have been largely investigated by both theoretical and experimental
approaches focused on their structure and interaction network. Many
studies have been also devoted to protonated water clusters, which
represent suitable and realistic models for understanding the proton
transfer mechanism in aqueous solutions.\citep{Qi_Bowman_protonatedwater_2017}
Furthermore, the interest in water clusters and systems where one
or more water molecules interact with other species has recently involved
methane and hydrogen clathrates,\citep{Conte_Bowman_Manybody_2015,Chen_Bowman_Methane-Water_2015,Homayoon_Bowman_H2-H2O_2015}
HCl hydrates,\citep{mancini_Bowman_mixedwaterHClpes_2013,mancini_Bowman_mizedwaterHClclusters_2014,Samanta_Reisler_Mixedclusters_2016}
or even solvated ions.\citep{kamarchik_Bowman_PEShydraredflouride_2014,kamarchik_Bowman_PEShydratedsodium_2011,kamarchik_bowman_PESwatercloride_2010,wang_kamarchik_PEShydratedNaClNaF_2016}

Focusing on homogeneous water clusters, experimental investigations
involving systems of different size ranging from the dimer up to the
decamer have shown that the vibrational spectral features of the OH
bonds are extremely sensitive to hydrogen interactions and dependent
on the specific cluster network,\citep{paul_okeefe_IRsmallclusters_1997,bouteiller_perchard_waterdimerspectra_2004,buck_huisken_waterclustersreview_2000}
while deuteration studies have demonstrated that the OH vibrational
frequencies may serve as a probe for hydrogen bonding.\citep{paul_saykally_ODstretching_1998}
Experimental findings also include the evidence that the OH stretches
involved in the hydrogen bonds undergo a red shift sometimes as large
as 600-700 cm\textsuperscript{-1},\citep{lakshminarayanan_ghosh_H2O45cluster_2005,steinbach_buch_OHhexamer_2004}
while the frequency of vibration of the free OH stretches is almost
unchanged and the bending region characterized by a slight but progressive
blue shift with increasing cluster size.\citep{paul_saykally_bendingwaterclusters_1999}
The main vibrational features of these clusters are distributed in
the 1500-4000 cm\textsuperscript{-1} region. The lowest in frequency
(at about 1600 cm\textsuperscript{-1}) can be assigned to the bending
motion,\citep{hirabayashi_yamada_kriptonxenonwaterclusters_2005}
while the other features are associated to the OH stretch and situated
at around 3000, 3500, and 3700 cm\textsuperscript{-1}.\citep{buck_sadley_spectraoctamer_1998}

From a theoretical point of view, important and pioneering work has
been performed by Xantheas and co-workers, who revealed that the potential
energy surface (PES) of even small clusters is very complicated because
of the many minima that can be located on it. They also showed that
for properly studying these systems a high level of electronic structure
theory must be employed, and that the zero-point-energy correction
is definitely not negligible to determine the relative stability between
the several minima.\citep{yoo_xantheas_newglobalminimum_2010} Energy
differences among these minima are often smaller than a fraction of
kcal/mol, so an investigation based exclusively on the global minimum
is probably not accurate enough to properly account for the overall
properties.\citep{bulusu_xiao_H2On11-13_2006,xantheas_dunningjr_structuretrimer_1993,maheshwary_gadre_structurelargewaterclusters_2001,wales_hodges_structurewaterclystersnless21_1998,xantheas_harrison_MP2structurewaterclusters_2002,xantheas_xantheas_abinitiowaterclusters_1995} 

The complexity of the water cluster PESs makes their construction
difficult, while the high level of electronic theory required rules
out the possibility to resort to on-the-fly \textit{ab initio} approaches.
However, the development in the past years of precise fitting procedures
has opened up the way to many theoretical investigations of variously-sized
water clusters.\citep{bowman_xie_somepes_2010,wang_paesani_hexamerisomers_2012,liu_bowman_LMMice_2012,ch_reisler_energytranfdimer_2012,wang_bowman_LMMhexamer_2013,babin_paesani_cagevsprism_2013,medders_paesani_manybodyelectrostatic_2013,liu_bowman_LMMliquidwater_2015,richardson_quntumtunnelingprism_2016}
Among them we recall the work done by Partridge and Schwenke\citep{partridge_Schwenke_PESH2Omonomer_1997}
in which they developed an accurate one body potential, the parametrization
of the water model by Burnham and Leslie,\citep{burnham_leslie_parametrizationwaterpes_1999}
or, more recently, the effort profused by the groups of Xantheas and
Bowman which led to more and more accurate water potentials.\citep{burnham_xantheas_addmonomer_TTM2--R_2002,fanourgakis_xantheas_reparam_TTM2--F_2006,huang_vanderavoird_HBB_2008,fanourgakis_xantheas_reparamTTM3--F_2008,shank_bowman_refitHBB_2009,Wang_Bowman_local-monomer_2011,wang_bowman_whbbpes_2011,wang_bowman_WHBB2_2016} 

A practical way to describe the PES of a water cluster is through
a many-body representation. Several studies, including those by Xantheas,
Clary, Paesani and their co-workers to name a few remarkable ones,
demonstrated that the representation can be truncated to the three-body
terms without significant loss of accuracy.\citep{xantheas_xantheas_manybodyinteractions_1994,gregory_clary_threebodywatertrimer_1995,medders_paesani_manybodyelectrostatic_2013}
In particular, Bowman's HBB,\citep{huang_vanderavoird_HBB_2008} WHBB,\citep{wang_bowman_WHBBpes_2009,wang_bowman_WHBB2_2016}
and WHBB2 PESs\citep{wang_bowman_WHBB2_2016} include terms up to
the three-body interaction and were shown to be very accurate and
flexible for water clusters of any size, thus permitting state-of-art
VCI calculations for the vibrational frequencies.\citep{Wang_Bowman_local-monomer_2011}
These calculations were based on the local monomer approach which
permits to accurately simulate spectra and to deal with even large
water clusters which otherwise would not be computationally affordable. 

We deem that an alternative, quantum dynamical theoretical approach
for spectroscopy calculations of water clusters could more realistically
describe the hydrogen bonding among water monomers and better uncover
possible resonances between overtones and fundamental vibrations.
The latter may for instance involve the OH bending overtones and the
OH stretching fundamental vibrations, which occur at nearby frequency
values. Such a quantum dynamical approach can be provided by the semiclassical
theory (SC) in its initial value representation version (SCIVR). SCIVR
builds a bridge between classical and quantum physics, since it allows
to approximate the quantum propagator reliably by using only dynamical
quantities that are generated from a classical simulation.\citep{Heller_SCspectroscopy_1981,shao_makri_backfrowardnopref_1999,Miller_Addingquantumtoclassical_2001,Shalashilin_Child_CCS_2004,Pollak_Perturbationseries_2007,Tatchen_Pollak_Onthefly_2009,Conte_Pollak_ThawedGaussian_2010,Conte_Pollak_ContinuumLimit_2012,Wehrle_Vanicek_NH3_2015,Antipov_Nandini_Mixedqcl_2015,Cina_Chapman_2011_smallmolecules,Nakamura_Ohta_SCDevelopment_2016,bonnet_rayez_Gaussweighting_2004}
Specifically, the time averaged version of the quantum propagator
is able to detect quantum effects on small- and medium-sized molecules
accurately.\citep{Kaledin_Miller_Timeaveraging_2003,Kaledin_Miller_TAmolecules_2003,DiLiberto_Ceotto_Prefactors_2016,Conte_Ceotto_NH3_2013,Ceotto_AspuruGuzik_Curseofdimensionality_2011,Ceotto_AspuruGuzik_Firstprinciples_2011,Ceotto_AspuruGuzik_Multiplecoherent_2009,Ceotto_AspuruGuzik_PCCPFirstprinciples_2009,Ceotto_Tantardini_Copper100_2010,Tamascelli_Ceotto_GPU_2014,Buchholz_Ceotto_MixedSC_2016,Buchholz_Ceotto_applicationMixed_2017,Zhuang_Ceotto_Hessianapprox_2012,Ceotto_Hase_AcceleratedSC_2013}
Recently, we have proposed a method called Divide-and-Conquer Semiclassical
Initial Value Representation (DC SCIVR)\citep{ceotto_conte_DCSCIVR_2017,DiLiberto_Ceotto_Jacobiano_2018}
which makes semiclassical dynamics viable also for large molecules.
In DC SCIVR the full-dimensional problem is divided into a set of
lower dimensional ones before proceeding with a proper set of SC calculations
constrained within the low-dimensional subspaces but still based on
the full-dimensional classical trajectory.

In this work we present a theoretical investigation of variously sized
water clusters by means of our recently established DC SCIVR. Results
show that quantum anharmonic effects are not negligible and dynamical
effects associated to the strong hydrogen-bond interactions are relevant.
We also illustrate a methodology for selecting a few relevant minima
in order to run semiclassical simulations with very reduced computational
costs yet retaining good accuracy. In the next Section we describe
the computational approach employed. Then we report our results starting
from the investigation of the vibrational features of the smallest
water clusters: the dimer and the trimer. Finally the focus shifts
to the water hexamer prism, for which we present some evidences of
the important role of hydrogen bond interactions, and to the water
decamer, whose vibrational features we are able to precisely describe
by employing just a few selected trajectories. The paper ends with
a brief summary and the presentation of our conclusions.

\section{Computational Methods\label{sec:Theory}}

The global PES employed in the calculations has been constructed according
to the many-body representation (truncated at the three-body level)
reported in Eq. (\ref{eq:Many-body}) 
\begin{center}
\begin{equation}
V({\bf q})=\sum_{i}^{N}V^{W}({\bf q}_{i})+\sum_{j>i}^{N}V^{W-W}({\bf q}_{i},{\bf q}_{j})+\sum_{k>j>i}^{N}V^{W-W-W}({\bf q}_{i},{\bf q}_{j},{\bf q}_{k}),\label{eq:Many-body}
\end{equation}
\par\end{center}

\noindent where the $W$ superscripts stand for ``water'', $N$
indicates the number of water monomers in the cluster, ${\bf q}$
represents the collection of all atomic coordinates, while ${\bf q}_{i}$
is the set of coordinates corresponding to atoms exclusively belonging
to the i-th water monomer. Specifically, we used the Partridge-Schwenke
potential for the 1-body ($W$) term;\citep{partridge_Schwenke_PESH2Omonomer_1997}
the 2-body (\textbf{$W-W$}) interaction surface has been extracted
from the highly accurate WHBB potential;\citep{wang_bowman_whbbpes_2011}
a new, computationally-efficient potential was built for the 3-body
($W-W-W)$ interaction starting from the same database of about 40,000
ab initio points used for the WHBB 3-body potential but employing
a previously developed fitting procedure for many-body interaction
potentials.\citep{Conte_Bowman_Manybody_2015,Conte_Bowman_Communication_2014,Chen_Bowman_Methane-Water_2015,Houston_Bowman_ModelFinal_2015,Conte_Bowman_CollisionsCH4-H2O_2015,Houston_Bowman_RoamingH2CO_2016}
This new 3-body potential is based on 1,181 polynomials, it has an
rms of 46 cm\textsuperscript{-1} calculated with respect to the database
of ab initio points (51 cm\textsuperscript{-1} for WHBB with maximum
polynomial order 5), and it is about 70 times faster than the 3-body
potential (maximum polynomial order 5) included in the WHBB suite.
A $W-W-W$ potential very similar in speed to the one employed here
has been recently and independently developed by Bowman and coworkers
and included in a new version of their water potential, named WHBB2.\citep{wang_bowman_WHBB2_2016}

Vibrational frequencies have been determined upon calculation of semiclassical
power spectra. Semiclassical approaches aim at regaining quantum information
starting from classically-evolved trajectories and their mathematical
formalism is obtained by approximating Feynman's quantum propagator.
Feynman's path integral formulation\citep{feynman_pathintegral_1965}
is a renowned way to represent the quantum propagator in which the
probability of going from an initial state $\text{\ensuremath{\mathbf{q}_{0}}}$
to a final one $\text{\ensuremath{\mathbf{q}_{t}}}$ can be obtained
by summing up over all the paths that connect the two states. A weight
that depends on its action is associated to each path
\begin{equation}
\text{\ensuremath{\left\langle \mathbf{q}_{t}\left|e^{-i\hat{H}t/\hbar}\right|\mathbf{q}_{0}\right\rangle }=\ensuremath{\int_{\mathbf{q}_{0}}^{\mathbf{q}_{t}}\mathcal{D}\left[\mathbf{q}\left(t\right)\right]e^{iS_{t}\left[\mathbf{q}_{0},\mathbf{q}_{t}\right]/\hbar}}}\label{eq:path_integral_propagator}
\end{equation}
Upon stationary-phase approximation, the integral in Eq. (\ref{eq:path_integral_propagator})
becomes a sum over the paths that give the major contribution. Such
paths are the classical ones and the approximation is exact up to
quadratic potentials.\citep{Tannor_book_2007} The result coincides
with the van Vleck version of the semiclassical propagator,\citep{VanVleck_SCpropagator_1928}
which is reported in Eq. \ref{eq:Van-Vleck_propagator}.
\[
\text{\ensuremath{\left\langle \mathbf{q}_{t}\left|e^{-i\hat{H}t/\hbar}\right|\mathbf{q}_{0}\right\rangle \approx\sum_{\text{cl. paths}}\sqrt{\frac{1}{\left(2\pi i\hbar\right)^{F}}\left|-\frac{\partial^{2}S_{t}^{cl}\left(\mathbf{q}_{0},\mathbf{q}_{t}\right)}{\partial\mathbf{q}_{t}\partial\mathbf{q}_{0}}\right|}e^{iS_{t}^{cl}\left(\mathbf{q}_{0},\mathbf{q}_{t}\right)/\hbar-i\nu\pi/2}}}
\]
\begin{equation}
=\text{\ensuremath{\sum_{\text{cl. paths}}\sqrt{\frac{1}{\left(2\pi i\hbar\right)^{F}}\left|\frac{\partial\mathbf{q}_{t}}{\partial\mathbf{p}_{0}}\right|^{-1}}e^{iS_{t}^{cl}\left(\mathbf{q}_{0},\mathbf{q}_{t}\right)/\hbar-i\nu\pi/2}}},\label{eq:Van-Vleck_propagator}
\end{equation}

\noindent where $F$ is the number of degrees of freedom, $\mathbf{p}_{0}$
is the initial momentum of the classical path, and $\upsilon$ is
the Maslov index, which ensures the continuity of the complex square
root. A more user-friendly version of the SC propagator has been worked
out by Miller, who introduced the Initial Value Representation (IVR)\citep{Miller_S-Matrix_1970}
of the van Vleck propagator. Application of this SCIVR propagator
to the survival amplitude of a generic reference wavefunction $\left|\chi\right\rangle $
leads to
\begin{equation}
\text{\ensuremath{\left\langle \chi\left|e^{-i\hat{H}t/\hbar}\right|\chi\right\rangle \approx\int\int}d\ensuremath{\mathbf{p}_{0}}d\ensuremath{\mathbf{q}_{0}\sqrt{\frac{1}{\left(2\pi i\hbar\right)^{F}}\left|\frac{\partial\mathbf{q}_{t}}{\partial\mathbf{p}_{0}}\right|}\chi^{*}\left(\mathbf{q}_{t}\right)\chi\left(\mathbf{q}_{0}\right)e^{iS_{t}^{cl}\left(\mathbf{p}_{0},\mathbf{q}_{0}\right)/\hbar-i\nu\pi/2}}}\label{eq:IVR_propagator}
\end{equation}
 Eq. (\ref{eq:IVR_propagator}) points out the two main advantages
of an IVR approach, i.e. the difficult quest for a solution to a double-boundary
problem is substituted by the easy generation of the classical trajectory
starting from its initial phase-space conditions ($\mathbf{p}_{0},\mathbf{q}_{0}$),
plus the removal of the partial derivative at the denominator of Eq.
(\ref{eq:Van-Vleck_propagator}) which may lead to an unphysical divergence
of the propagator. Another milestone contribution to SC dynamics came
from Heller with the introduction of the coherent state representation.
Coherent states are suitable to describe bound as well as unbound
systems since their projection onto the coordinate space consists
in a Gaussian-shaped real part and a free-particle imaginary part
as follows\citep{Heller_FrozenGaussian_1981,Heller_Cellulardynamics_1991}
\begin{equation}
\text{\ensuremath{\left\langle \mathbf{x}\left|\mathbf{p}_{t}\mathbf{q}_{t}\right.\right\rangle }=\ensuremath{\left(\frac{det(\Gamma)}{\pi^{F}}\right)^{\frac{1}{4}}e^{-\frac{1}{2}\left(\mathbf{x}-\mathbf{q}_{t}\right)^{T}\Gamma\left(\mathbf{x}-\mathbf{q}_{t}\right)+\frac{i}{\hbar}\mathbf{p}_{t}^{T}\left(\mathbf{x}-\mathbf{q}_{t}\right)}}}\label{eq:Coherent_xspace}
\end{equation}
 The width of the multidimensional coherent state is generally chosen
to be a diagonal matrix ($\Gamma$). In the case of vibrational studies
the width matrix can be built by setting its diagonal elements equal
to the square roots of the eigenvalues of the Hessian matrix at the
equilibrium geometry, \textit{i.e.} the harmonic frequencies. The
semiclassical propagator in the coherent state representation is the
Herman Kluk (HK) propagator\citep{Herman_Kluk_SCnonspreading_1984}
\begin{equation}
\text{\ensuremath{\left\langle \chi\left|e^{-i\hat{H}t/\hbar}\right|\chi\right\rangle \approx\left(\frac{1}{2\pi\hbar}\right)^{F}\iintop}d\ensuremath{\mathbf{p}_{0}}d\ensuremath{\mathbf{q}_{0}C_{t}\left(\mathbf{p}_{0},\mathbf{q}_{0}\right)e^{\frac{i}{\hbar}S_{t}\left(\mathbf{p}_{0},\mathbf{q}_{0}\right)}\left\langle \chi\right.\left|\mathbf{p}_{t}\mathbf{q}_{t}\left\rangle \right\langle \mathbf{p}_{0}\mathbf{q}_{0}\right|\left.\chi\right\rangle },}\label{eq:HHKK_propagator}
\end{equation}
where the pre-exponential factor - $C_{t}\left(\mathbf{p}_{0},\mathbf{q}_{0}\right)$
- accounts for quantum effects but is affected by the possible chaotic
behavior of the classical trajectories initiated from the phase space
points $\text{\ensuremath{\left(\mathbf{p}_{0},\mathbf{q}_{0}\right)}}$.
\begin{equation}
\text{\ensuremath{C_{t}\left(\mathbf{p}_{0},\mathbf{q}_{0}\right)}=\ensuremath{\sqrt{\mbox{det}\left[\frac{1}{2}\left(\mathbf{M_{qq}}+{\bf \Gamma}^{-1}\mathbf{M_{pp}}\mathbf{\Gamma}+\frac{i}{\hbar{\bf \Gamma}}\mathbf{M_{pq}}-i\mathbf{\hbar{\bf \Gamma}M_{qp}}\right)\right]}}},\label{eq:HK_prefactor}
\end{equation}

\noindent where ${\bf M_{ij}}=\partial{\bf i}_{t}/\partial{\bf j}_{0}\quad{\bf i},{\bf j}={\bf p},{\bf q}$
is itself a matrix which represents a generic element of the monodromy
matrix.\citep{Tannor_book_2007,Wang_Miller_GeneralizedFilinov_2001} 

The power spectrum of the Hamiltonian $\text{\ensuremath{\hat{H}}}$
is the Fourier transform of Eq. (\ref{eq:HHKK_propagator}), \emph{i.e.}
\begin{equation}
\text{I\ensuremath{\left(E\right)}=\ensuremath{\left(\frac{1}{2\pi\hbar}\right)^{F+1}\int_{-\infty}^{+\infty}}dt \ensuremath{e^{iEt/\hbar}\iintop}d\ensuremath{\mathbf{p}_{0}}d\ensuremath{\mathbf{q}_{0}C_{t}\left(\mathbf{p}_{0},\mathbf{q}_{0}\right)e^{\frac{i}{\hbar}S_{t}\left(\mathbf{p}_{0},\mathbf{q}_{0}\right)}\left\langle \chi\right.\left|\mathbf{p}_{t}\mathbf{q}_{t}\left\rangle \right\langle \mathbf{p}_{0}\mathbf{q}_{0}\right|\left.\chi\right\rangle }.}\label{eq:HHKK_spectrum}
\end{equation}
Unfortunately the standard formulation of the HK propagator is difficult
to converge and computationally very demanding.\citep{Kay_Multidim_1994,Kay_Numerical_1994}
To overcome this issue, Kaledin and Miller demonstrated that it is
possible to time-average (TA) Eq. (\ref{eq:HHKK_propagator}) to arrive
to an expression for the spectral density $\text{I(E)}$ - see Eq.
(\ref{eq:TA_spectrum}) - where the phase-space integrand is positive-definite
and, consequently, the integral is much easier to converge\citep{Kaledin_Miller_Timeaveraging_2003}
\begin{equation}
\text{I\ensuremath{\left(E\right)}=\ensuremath{\left(\frac{1}{2\pi\hbar}\right)^{F}\iintop}d\ensuremath{\mathbf{p}_{0}}d\ensuremath{\mathbf{q}_{0}\frac{1}{2\pi\hbar T}\left|\intop_{0}^{T}dte^{\frac{i}{\hbar}\left[S_{t}\left(\mathbf{p}_{0},\mathbf{q}_{0}\right)+Et+\phi\left(t\right)\right]}\left\langle \boldsymbol{\chi}\left|\mathbf{p}_{t}\mathbf{q}_{t}\right.\right\rangle \right|^{2}}.}\label{eq:TA_spectrum}
\end{equation}
 Eq (\ref{eq:TA_spectrum}) is very accurate for small and medium
sized molecules, but, unfortunately, runs out of steam when the system
dimensionality gets higher than 25-30 degrees of freedom due to the
so-called curse of dimensionality. To overcome this issue we have
recently developed a Divide-and-Conquer semiclassical method, which
allows to obtain the overall power spectrum as a composition of partial
spectra.\citep{ceotto_conte_DCSCIVR_2017} 

In the following, we briefly recall how DC SCIVR works. The basic
idea is to compute a set of power spectra by operating in lower-dimensional
subspaces but keeping the dynamics full-dimensional. Among the approaches
we have recently illustrated for an effective grouping of the modes
into different subspaces,\citep{DiLiberto_Ceotto_Jacobiano_2018}
in this work we have adopted the so-called Hessian approach, which
consists in averaging the Hessian off-diagonal elements $\tilde{H}_{ij}$
along a preliminary trajectory with harmonic zero-point energy and
in comparing them to a threshold value $\varepsilon$. If the vibrational
modes i and j satisfy the condition $\tilde{H}_{ij}\geq\varepsilon$,
they are enrolled in the same subspace. Also, they belong to the same
subspace if a third mode k exists such that the conditions $\tilde{H}_{ik}\geq\varepsilon$
and $\tilde{H}_{kj}\geq\varepsilon$ are both satisfied. Eq. (\ref{eq:TA_spectrum})
consequently changes to 
\begin{equation}
\text{\ensuremath{\widetilde{I}\left(E\right)}=\ensuremath{\left(\frac{1}{2\pi\hbar}\right)^{M}\iintop}d\ensuremath{\tilde{\mathbf{p}}\left(0\right)}d\ensuremath{\tilde{\mathbf{q}}\left(0\right)\frac{1}{2\pi\hbar T}\left|\intop_{0}^{T}e^{\frac{i}{\hbar}\left[\tilde{S}_{t}\left(\tilde{\mathbf{p}}\left(0\right),\tilde{\mathbf{q}}\left(0\right)\right)+Et+\tilde{\phi}_{t}\right]}\left\langle \tilde{\boldsymbol{\chi}}\left|\tilde{\mathbf{p}}\left(t\right)\tilde{\mathbf{q}}\left(t\right)\right.\right\rangle dt\right|^{2}},}\label{eq:projected_spectra}
\end{equation}
\noindent where $\tilde{f}$ indicates the projection of the generic
$f$ quantity onto an M-dimensional subspace. Matrices (Hessian and
Gaussian width ones) as well as vectors (momentum and position) can
be easily projected by means of Hinsen and Kneller's singular value
decomposition.\citep{Hinsen_Kneller_SingValueDecomp_2000,Harland_Roy_SCIVRconstrained_2003,ceotto_conte_DCSCIVR_2017}
Projection of the action is more elaborated due to the general non-separability
of the potential energy. In fact, the projected action is calculated
through straightforward projection of the kinetic energy, which is
separable, and by means of a projected potential - see Eq. (\ref{eq:Proj_Potential}).
In the projected potential the variables external to the M-dimensional
subspace (${\bf q}_{F-M})$ are treated as parameters and a time-dependent
field ($\lambda$), able to account for the non-separability of the
potential and to regain the exact potential in separable instances,
is introduced.\citep{ceotto_conte_DCSCIVR_2017,DiLiberto_Ceotto_Jacobiano_2018}

\begin{equation}
\begin{array}{c}
V({\bf \tilde{{\bf q}}}_{M})=V({\bf \tilde{{\bf q}}}_{M};{\bf q}_{F-M}^{eq})+\lambda(t);\\
\lambda(t)=V({\bf \tilde{{\bf q}}}_{M};{\bf q}_{F-M})-V({\bf \tilde{{\bf q}}}_{M}^{eq};{\bf q}_{F-M})-V({\bf \tilde{{\bf q}}}_{M};{\bf q}_{F-M}^{eq})
\end{array}\label{eq:Proj_Potential}
\end{equation}

For water clusters, due to the large number of low frequency modes
which may make spectra very noisy, it is necessary to introduce an
additional device consisting in giving no initial kinetic energy to
modes different from bendings and OH stretchings. Several of these
modes present very low vibrational features associated to the libration
motions of frustrated translations and rotations. Below we will show
that this \textit{ad-hoc} approximation does not spoil the accuracy
of the calculated frequencies for bendings and stretchings. Furthermore,
we employed a recently developed and accurate approximation to the
pre-exponential factor,\citep{DiLiberto_Ceotto_Prefactors_2016} which
has permitted to retain the chaotic trajectories that in a basic application
of DC SCIVR would have been otherwise discarded. The approximation
exploits the exact Log-derivative formulation of the pre-exponential
factor\citep{Gelabert_Miller_logderivative_2000} in which the latter
depends on a matrix $\mathbf{R}_{t}$ determined by solving an appropriate
Riccati equation 
\begin{equation}
\text{\text{\ensuremath{C_{t}\left(\mathbf{p}_{0},\mathbf{q}_{0}\right)}=}\ensuremath{\sqrt{\mbox{det}\left[\frac{1}{2}\left(I+\frac{i}{\hbar{\bf \Gamma}}\mathbf{R}_{t}\right)\right]}e^{\frac{1}{2}\intop_{0}^{t}d\tau\text{Tr}\left[\mathbf{R}_{\tau}\right]}}},
\end{equation}
where $\mathbf{R}_{t}$ can be approximated as 
\begin{equation}
\text{\ensuremath{\mathbf{R}_{t}}=-\ensuremath{\frac{i}{2}\left[\frac{\mathbf{K}_{t}}{\hbar\boldsymbol{\Gamma}}+\hbar\boldsymbol{\Gamma}\right]}+\ensuremath{\frac{i}{4}\frac{\left(\hbar\boldsymbol{\Gamma}-\frac{\mathbf{K}_{t}}{\hbar\boldsymbol{\Gamma}}\right)^{2}}{\left(\hbar\boldsymbol{\Gamma}+\frac{\mathbf{K}_{t}}{\hbar\boldsymbol{\Gamma}}\right)}},}
\end{equation}

\noindent and $\text{\ensuremath{\mathbf{K}_{t}}}$ is the Hessian
matrix. 

According to the convergence pattern of TA-SCIVR, the power spectra
would require generation of a number of trajectories of the order
of a thousand per degree of freedom to reach convergence. A reliable
procedure to alleviate computational overheads is needed for application
of DC SCIVR to large molecular systems. For this purpose, we implemented
the Multiple Coherent State (MC) approach into DC-SCIVR by running
one trajectory per each of the $N_{st}$ coherent states that make
up the reference state $|\chi\rangle$ according to Eq. (\ref{eq:MC_reference_state}). 

\begin{equation}
|\chi\rangle=\sum_{k=1}^{N_{st}}\prod_{j=1}^{M}\text{\ensuremath{\left|p_{eq,j}^{(k)}q_{eq,j}^{(k)}\right\rangle }+\ensuremath{\xi_{j}^{(k)}\left|-p_{eq,j}^{(k)}q_{eq,j}^{(k)}\right\rangle }},\label{eq:MC_reference_state}
\end{equation}
\noindent where $\xi_{j}^{(k)}$ is a parameter that allows to distinguish
between different vibrational signals according to their symmetry
or parity. In this way, as demonstrated in the literature,\citep{Ceotto_AspuruGuzik_Multiplecoherent_2009,Ceotto_AspuruGuzik_Curseofdimensionality_2011,Conte_Ceotto_NH3_2013,Gabas_Ceotto_Glycine_2017}
accurate spectra can be recovered by running just a few or even a
single classical trajectory provided it is close in energy to the
actual quantum vibrational frequency. To get more information than
a traditional discrete Fourier transform, the time integration of
Eq.(\ref{eq:projected_spectra}) is performed using the compress sensing
signal processing technique.\citep{Markovich_Aspuru_CompressSensing_2016}

The divide-and-conquer approach can be also implemented to calculate
classical-like spectral densities from the Fourier transform of the
velocity-velocity correlation function 
\begin{equation}
I\left(E\right)=\int dte^{iEt}\biggl\langle\mathbf{v}\left(t\right)\mathbf{v}\left(0\right)\biggr\rangle=\int_{-\infty}^{+\infty}dte^{iEt}\int d\mathbf{q}_{0}d\mathbf{p}_{0}\rho\left(\mathbf{q}_{0},\mathbf{p}_{0}\right)\mathbf{v}\left(t\right)\mathbf{v}\left(0\right)\label{eq:Classical_spectrum}
\end{equation}
By adding a further integration ($\frac{1}{T}\int_{0}^{T}dt$) we
can derive the time-averaged version of Eq. (\ref{eq:Classical_spectrum}),
similarly to what Miller and co-workers have obtained for semiclassical
spectral densities,\citep{Kaledin_Miller_Timeaveraging_2003,Kaledin_Miller_TAmolecules_2003}
and Kaledin and Bowman for classical simulations \citep{Kaledin_Bowman_classical_2004}
 
\begin{equation}
I\left(E\right)=lim_{T\rightarrow+\infty}\int d\mathbf{q}_{0}d\mathbf{p}_{0}\rho\left(\mathbf{p}_{0},\mathbf{q}_{0}\right)\frac{1}{2T}\Biggl|\int_{0}^{T}dte^{iE\left(t\right)}\mathbf{v}\left(t\right)\Biggr|^{2}.
\end{equation}
Reduced dimensional spectra can be calculated by means of the projected
classical quantities obtained with the same procedure employed in
DCSCIVR.\citep{ceotto_conte_DCSCIVR_2017,DiLiberto_Ceotto_Jacobiano_2018}
The final working formula is
\begin{equation}
I\left(E\right)=lim_{T\rightarrow+\infty}\int d\mathbf{\tilde{q}}_{0}d\mathbf{\tilde{p}}_{0}\tilde{\rho}\left(\tilde{\mathbf{q}}_{0},\mathbf{\tilde{p}}_{0}\right)\frac{1}{2T}\Biggl|\int_{0}^{T}dte^{iE\left(t\right)}\mathbf{\tilde{v}}\left(t\right)\Biggr|^{2},\label{eq:classical_projected}
\end{equation}

\noindent where $\tilde{\rho}({\bf p}_0,{\bf q}_0)$ is the sampling
phase-space distribution function in reduced dimensionality.

\section{Results and Discussion\label{sec:Results}}

We start off by demonstrating the reliability of our DC-SCIVR calculations
of small water clusters such as the dimer and trimer, for which accurate
MultiMode and experimental energy levels are available. Then, we move
to the water hexamer showing the influence of dynamical effects on
spectral densities. Finally, as an application to a larger system,
we calculate the vibrational energy levels of the water decamer.

\subsection{Water Dimer (H\protect\textsubscript{2}O)\protect\textsubscript{2}}

\begin{figure}
\begin{centering}
\includegraphics[scale=0.4]{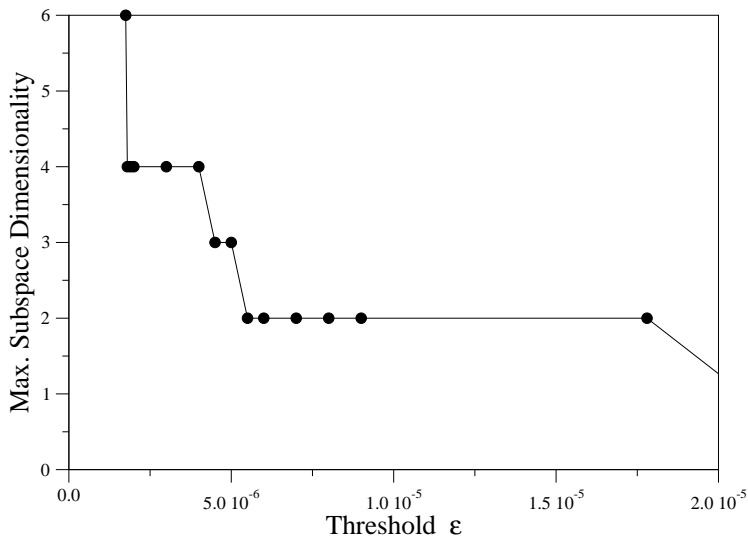}
\par\end{centering}
\centering{}\caption{\label{fig:Threshold_dimer}Dependence of the maximum subspace dimensionality
on the arbitrary Hessian threshold for the water dimer.}
\end{figure}

The smallest water cluster - the dimer - has 12 vibrational degrees
of freedom, two of which are bendings while other four are OH stretchings.
To calculate these six relevant frequencies of vibration, we first
tried to employ a 6-dimensional work subspace to collect all bendings
and stretchings together but, unfortunately, it was not possible to
get a well-resolved spectroscopic signal. To overcome this issue,
we applied the DC-SCIVR partitioning to decrease the maximum subspace
dimension down to 2 as suggested by the large plateau which can be
easily identified from the analysis of the maximum subspace dimensionality
vs Hessian threshold dependence shown in Figure (\ref{fig:Threshold_dimer}).

In our calculations we employed a threshold value $\varepsilon=1.8\cdot10^{-5}$.
Table (\ref{tab:dimer}) presents experimental values,\citep{bouteiller_perchard_waterdimerspectra_2004}
MultiMode (MM) and Local Monomer Model (LMM) data,\citep{Wang_Bowman_local-monomer_2011}
and our semiclassical DC-SCIVR results based on different sets of
trajectories. The outcomes of single-trajectory simulations (based
on a dynamics 30,000 atomic time units long) that employed the multiple
coherent procedure within the subspaces (MC-DC SCIVR) are reported
in the last column of the Table. The mean absolute error (MAE) with
respect to experimental values (78 cm\textsuperscript{-1}) is not
satisfactory especially if compared to the MAE values of the benchmark
MM and LMM calculations (25 and 23 cm\textsuperscript{-1} respectively).
To understand the reasons of such inaccuracy and try to improve results,
we investigated the presence of additional minima on the surface which
may be neglected in a single-trajectory simulation but, at the same
time, may influence the semiclassical results. This is true if these
local minima are very close in energy to the global minimum.\citep{Gabas_Ceotto_Glycine_2017}
For this purpose we explored the PES by means of damped-dynamics runs,
and found 4 local minima just about 200 cm\textsuperscript{-1} higher
in energy than the global minimum. The damped dynamics was performed
by sampling 1,000 trajectories according to a Husimi distribution
around the global minimum. The damping parameter has been chosen according
to a trade-off between the necessity of exploring large regions of
the PES and the need to limit the simulation times. In practice, we
decreased the kinetic energy by a factor equal to 0.99 at each step
of the dynamics and checked that it was terminated in a minimum (either
local or global) by looking at the sign of the eigenvalues of the
Hessian matrix calculated at that geometry. Then, we re-applied the
MC-DC-SCIVR approach by running 5 trajectories per subspace this time,
with each trajectory starting from a different minimum and the MAE
shifts from 78 to a much improved 48 cm\textsuperscript{-1}, which
is not far from the accuracy obtained in other previous semiclassical
calculations.\citep{Conte_Ceotto_NH3_2013} This outcome will be exploited
in treating much larger clusters for which semiclassical calculations
can be performed only if based on a small number of trajectories.
The multiple-minima effect can also partially explain the difference
between DC-SCIVR simulations (that visit all 5 minima) and the single-well
reference MM or LMM calculations. 

To check the reliability of MC-DC-SCIVR simulations we also performed
standard, fully converged, DC-SCIVR simulations. Convergence has been
reached by employing 10,000 trajectories, but 5,000-trajectory simulations
were found to be already reliable. We started all the trajectories
with the cluster in its equilibrium geometry. Initial atomic velocities
were extracted, for each subspace calculation, from the chosen distribution
of the normal mode initial kinetic energy. Specifically, for modes
included in the subspace under investigation a Husimi distribution
centered on momentum values corresponding to one quantum of harmonic
excitation was employed; other bending and stretching motions belonging
to different subspaces were instead assigned the corresponding harmonic
zero-point energy contribution. Finally, as anticipated in the previous
Section, all other low frequency modes were given no initial kinetic
energy. Furthermore, harmonic frequencies served to define the coherent
state and Husimi distribution widths. Each trajectory was evolved
for a total of 30,000 atomic time units. Semiclassical investigations
performed on trajectories twice as long provided no significant differences
in the vibrational frequencies indicating that 30,000 atomic units
is a long enough evolution time to achieve numerical convergence.
MAE values for simulations based on 10,000 and 5,000 trajectories
are equal to 32 and 39 cm\textsuperscript{-1}respectively. The MAE
value obtained with MC-DC-SCIVR based on 5 tailored trajectories (48
cm\textsuperscript{-1}) is not far from those values, confirming
the reliability of the computationally-cheaper approach. From the
results it is evident the separation between the bending and stretching
frequencies, and that MC-DC SCIVR must be adopted upon identification
of all relevant minima. A final comparison to the experiment demonstrates
that standard DC SCIVR is also able to detect the high frequency stretching
overtones with reasonable accuracy.

\begin{figure}
\begin{centering}
\includegraphics[scale=0.5]{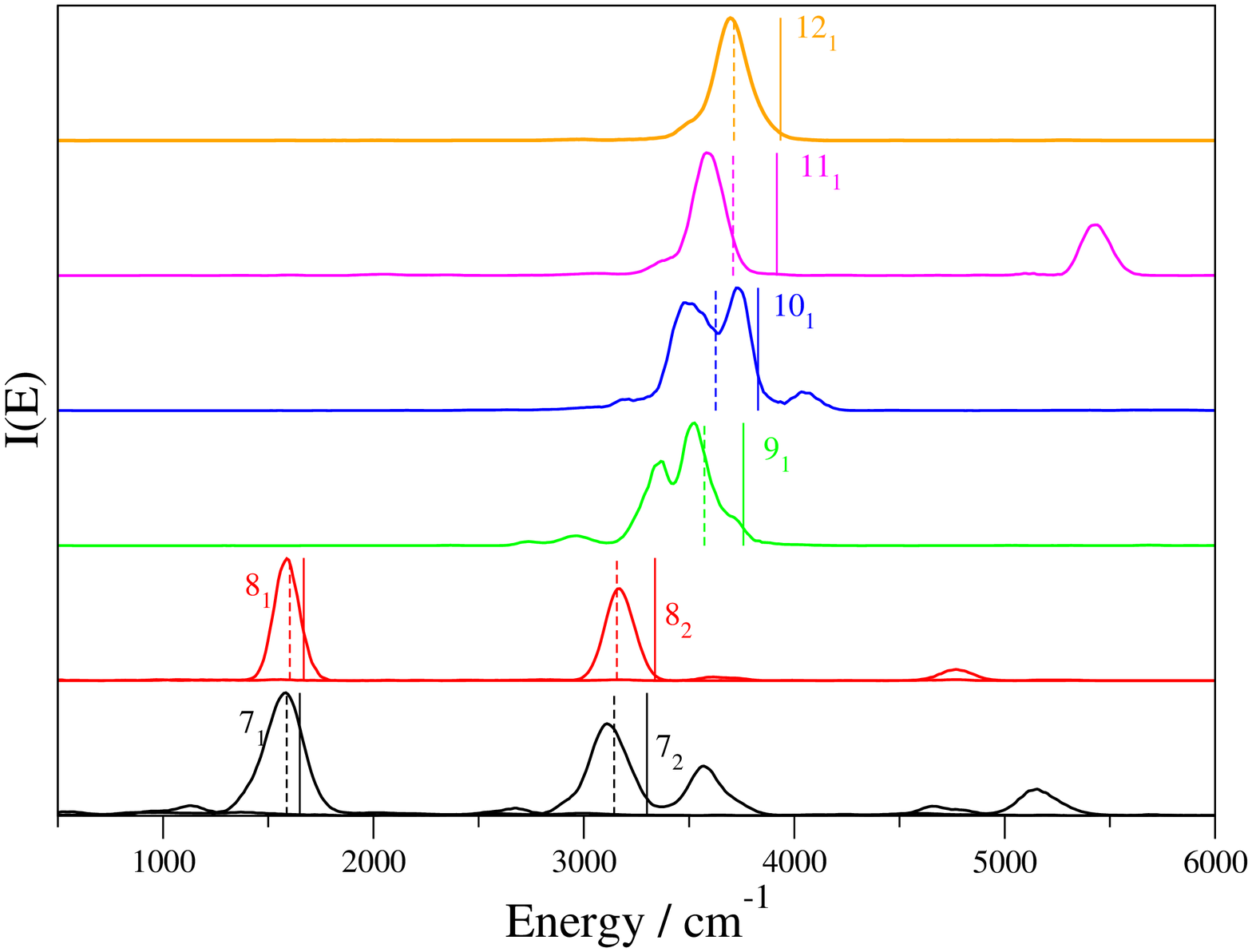}
\par\end{centering}
\caption{\label{fig:Vibrational-spectra-dimer}MC-DC-SCIVR vibrational spectra
of bendings and stretches of the water dimer based on 5 trajectories
per subspace. The vertical solid lines indicate the harmonic estimates,
while the dashed vertical lines the MultiMode results. The bending
fundamental and overtone signals were obtained by tuning the reference
state according to Eq. (\ref{eq:MC_reference_state}).}
\end{figure}

Figure (\ref{fig:Vibrational-spectra-dimer}) shows the peaks obtained
with MC-DC SCIVR employing 5 trajectories per subspace, and reports
MultiMode and harmonic frequency estimates for a visual comparison.
We observe that in our semiclassical spectra, because of the interaction
between different modes, multiple-peak features appear as, for instance,
in the case of mode 10 (which shows a shoulder at the frequency of
mode 9), or as in the case of mode 9 and overtones of modes 7 and
8. The interaction between the bending overtones and the lower frequency
stretches involved in hydrogen bondings is a general feature of water
clusters which we found also in larger ones and which complicates
the aspect of the simulated spectra. 

\begin{table}
\begin{centering}
\caption{\label{tab:dimer}Vibrational frequencies of the water dimer, in $\text{c\ensuremath{m^{-1}}}.$
Assignments of mode excitations are reported in the first column;
the following columns present, in order, the experimental results
(Exp), the harmonic estimates (HO), Multimode (MM) and Local Monomer
Model (LMM) results, DC-SCIVR frequencies obtained from 5,000 and
10,000 trajectories, MC-DC-SCIVR frequencies based on 5 trajectories,
and MC-DC-SCIVR results from a single trajectory started from the
global minimum. The mean absolute errors (MAE) are relative to the
experimental values. \protect\textsuperscript{a} from Ref. \citenum{bouteiller_perchard_waterdimerspectra_2004};
\protect\textsuperscript{b} from Ref. \citenum{Wang_Bowman_local-monomer_2011}.}
\par\end{centering}
\centering{}%
\begin{tabular}{ccccccccc}
\hline 
{\footnotesize{}Index} & {\footnotesize{}Exp\textsuperscript{a}} & {\footnotesize{}HO} & {\footnotesize{}MM}{\scriptsize{}\textsuperscript{b}} & {\footnotesize{}LMM\textsuperscript{b}} & {\footnotesize{}DC SCIVR\textsubscript{10k}} & {\footnotesize{}DC SCIVR$_{\text{5k}}$} & {\footnotesize{}MC-DC SCIVR\textsubscript{5 trajs, multmin}} & {\footnotesize{}MC-DC SCIVR$_{\text{1 traj}}$}\tabularnewline
\hline 
{\footnotesize{}$7_{1}$} & {\footnotesize{}1600} & {\footnotesize{}1650} & {\footnotesize{}1588} & {\footnotesize{}1595} & {\footnotesize{}1597} & {\footnotesize{}1597} & {\footnotesize{}1562} & {\footnotesize{}1572}\tabularnewline
\hline 
{\footnotesize{}$8_{1}$} & {\footnotesize{}1617} & {\footnotesize{}1669} & {\footnotesize{}1603} & {\footnotesize{}1602} & {\footnotesize{}1585} & {\footnotesize{}1578} & {\footnotesize{}1588} & {\footnotesize{}1578}\tabularnewline
\hline 
{\footnotesize{}$7_{2}$} & {\footnotesize{}3163} & {\footnotesize{}3300} & {\footnotesize{}3144} & {\footnotesize{}3153} & {\footnotesize{}3154} & {\footnotesize{}3178} & {\footnotesize{}3128} & {\footnotesize{}3156}\tabularnewline
\hline 
{\footnotesize{}$8_{2}$} & {\footnotesize{}3194} & {\footnotesize{}3338} & {\footnotesize{}3157} & {\footnotesize{}3168} & {\footnotesize{}3130} & {\footnotesize{}3100} & {\footnotesize{}3180} & {\footnotesize{}3156}\tabularnewline
\hline 
{\footnotesize{}$9_{1}$} & {\footnotesize{}3591} & {\footnotesize{}3758} & {\footnotesize{}3573} & {\footnotesize{}3550} & {\footnotesize{}3550} & {\footnotesize{}3539} & {\footnotesize{}3526} & {\footnotesize{}3356}\tabularnewline
\hline 
{\footnotesize{}$10_{1}$} & {\footnotesize{}3661} & {\footnotesize{}3828} & {\footnotesize{}3627} & {\footnotesize{}3637} & {\footnotesize{}3690} & {\footnotesize{}3693} & {\footnotesize{}3680} & {\footnotesize{}3540}\tabularnewline
\hline 
{\footnotesize{}$11_{1}$} & {\footnotesize{}3734} & {\footnotesize{}3917} & {\footnotesize{}3709} & {\footnotesize{}3701} & {\footnotesize{}3670} & {\footnotesize{}3671} & {\footnotesize{}3582} & {\footnotesize{}3628}\tabularnewline
\hline 
{\footnotesize{}$12_{1}$} & {\footnotesize{}3750} & {\footnotesize{}3935} & {\footnotesize{}3713} & {\footnotesize{}3724} & {\footnotesize{}3764} & {\footnotesize{}3764} & {\footnotesize{}3717} & {\footnotesize{}3690}\tabularnewline
\hline 
{\footnotesize{}MAE} & {\footnotesize{}-} & {\footnotesize{}136} & {\footnotesize{}25} & {\footnotesize{}23} & {\footnotesize{}32} & {\footnotesize{}39} & {\footnotesize{}48} & {\footnotesize{}78}\tabularnewline
\hline 
 & {\footnotesize{}7207} &  &  &  &  & {\footnotesize{}7266} &  & \tabularnewline
\hline 
 & {\footnotesize{}7362} &  &  &  &  & {\footnotesize{}7336} &  & \tabularnewline
\hline 
 & {\footnotesize{}5328} &  &  &  &  & {\footnotesize{}5375} &  & \tabularnewline
\hline 
\end{tabular}
\end{table}

\subsection{Water Trimer (H\protect\textsubscript{2}O)\protect\textsubscript{3} }

The water trimer is made of 21 vibrational degrees of freedom, 3 of
which are bendings and 6 are OH stretches. In order to reduce the
computational burden, we employed a version of the three body potential
which is, as anticipated, very fast to compute and retains quite well
the accuracy of the original WHBB 3-body potential. A first analysis
of the trimer is obtained by looking at its Hessian threshold trend.
We observe that employing the same threshold value used for the dimer
would lead to label all vibrational modes as indepedent ones, reflecting
the decrement in magnitude of the off-diagonal elements of the trimer
Hessian. Furthermore, for the trimer, three plateau can be clearly
identified at maximum subspace dimensionality values of 8, 6 and 1.
However, similarly to the dimer case, spectra projected onto high-dimensional
subspaces cannot be well resolved and the maximum subspace dimensionality
still consistent with a resolved spectrum is 3. For this reason and
to keep working with subspaces as high dimensional as possible, we
used a threshold value for the trimer equal to $1.5\cdot10^{-5}$.
The relevant 9-dimensional space has been consequently divided into
mono, bi- and tri-dimensional subspaces. In particular, modes 17,18,19
have been enrolled into a 3-dimensional subspace, while modes 16 and
20 into a bidimensional one. Generation of the initial conditions
and the subsequent dynamics have been performed according to the methodology
already described for the dimer.
\begin{table}
\centering{}\caption{\label{tab:trimer}Main fundamental vibrational frequencies of the
water trimer, in $\text{c\ensuremath{m^{-1}}}.$ Labels are the same
as in Table (\ref{tab:dimer}). Last column shows classical-like results
from Eq. (\ref{eq:classical_projected}) based on the Fourier transform
(FT) of the velocity-velocity autocorrelation function (C\protect\textsubscript{vv}).
MAE values are relative to MultiMode (MM) results.}
\begin{tabular}{cccccccc}
{\scriptsize{}Index} & {\scriptsize{}HO} & {\scriptsize{}MM\citep{Wang_Bowman_local-monomer_2011}} & {\scriptsize{}LMM\citep{Wang_Bowman_local-monomer_2011}} & {\scriptsize{}DC SCIVR$\text{\ensuremath{_{10k}}}$} & {\scriptsize{}MC-DC SCIVR$_{\text{10trajs,multmin}}$} & {\footnotesize{}MC-DC SCIVR$_{\text{1 traj}}$} & {\footnotesize{}$FT(\text{\ensuremath{C_{vv}}})$}\tabularnewline
\hline 
{\scriptsize{}$13_{1}$} & {\scriptsize{}1661} & {\scriptsize{}1597} & {\scriptsize{}1602} & {\scriptsize{}1584} & {\scriptsize{}1575} & {\scriptsize{}1534} & {\scriptsize{}1520}\tabularnewline
\hline 
{\scriptsize{}$14_{1}$} & {\scriptsize{}1664} & {\scriptsize{}1600} & {\scriptsize{}1614} & {\scriptsize{}1595} & {\scriptsize{}1637} & {\scriptsize{}1528} & {\scriptsize{}1520}\tabularnewline
\hline 
{\scriptsize{}$15_{1}$} & {\scriptsize{}1681} & {\scriptsize{}1623} & {\scriptsize{}1615} & {\scriptsize{}1627} & {\scriptsize{}1634} & {\scriptsize{}1530} & {\scriptsize{}1516}\tabularnewline
\hline 
{\scriptsize{}$16_{1}$} & {\scriptsize{}3664} & {\scriptsize{}3486} & {\scriptsize{}3489} & {\scriptsize{}3440} & {\scriptsize{}3386} & {\scriptsize{}3426} & {\scriptsize{}3536}\tabularnewline
\hline 
{\scriptsize{}$17_{1}$} & {\scriptsize{}3703} & {\scriptsize{}3504} & {\scriptsize{}3500} & {\scriptsize{}3450} & {\scriptsize{}3400} & {\scriptsize{}3547} & {\scriptsize{}3548}\tabularnewline
\hline 
{\scriptsize{}$18_{1}$} & {\scriptsize{}3711} & {\scriptsize{}3514} & {\scriptsize{}3510} & {\scriptsize{}3247} & {\scriptsize{}3380} & {\scriptsize{}3151} & {\scriptsize{}3480}\tabularnewline
\hline 
{\scriptsize{}$19_{1}$} & {\scriptsize{}3911} & {\scriptsize{}3709} & {\scriptsize{}3718} & {\scriptsize{}3640} & {\scriptsize{}3610} & {\scriptsize{}3706} & {\scriptsize{}3676}\tabularnewline
\hline 
{\scriptsize{}$20_{1}$} & {\scriptsize{}3916} & {\scriptsize{}3715} & {\scriptsize{}3718} & {\scriptsize{}3700} & {\scriptsize{}3675} & {\scriptsize{}3652} & {\scriptsize{}3697}\tabularnewline
\hline 
{\scriptsize{}$21_{1}$} & {\scriptsize{}3918} & {\scriptsize{}3720} & {\scriptsize{}3719} & {\scriptsize{}3736} & {\scriptsize{}3760} & {\scriptsize{}3684} & {\scriptsize{}3640}\tabularnewline
\hline 
{\scriptsize{}MAE} & {\scriptsize{}151} & - & {\scriptsize{}6} & {\scriptsize{}54} & {\scriptsize{}65} & {\scriptsize{}88} & {\scriptsize{}58}\tabularnewline
\end{tabular}
\end{table}
Table (\ref{tab:trimer}) shows our DC-SCIVR (based on 10,000 trajectories,
30,000 atomic time units long) and MC-DC-SCIVR results compared to
the MultiMode and Local Monomer Model ones.\citep{Wang_Bowman_local-monomer_2011}
Once again, a single trajectory is insufficient to recover the correct
semiclassical spectral features, and a preliminary exploration of
the Potential Energy Surface is required for application of MC-DC
SCIVR. We found nine different local minima very close in energy to
the global one. We repeated the same MC-DC-SCIVR procedure described
in the dimer section, running in this case 10 trajectories for each
subspace, each one centered on a different minimum (global or local). 

\begin{figure}
\begin{centering}
\includegraphics[scale=0.5]{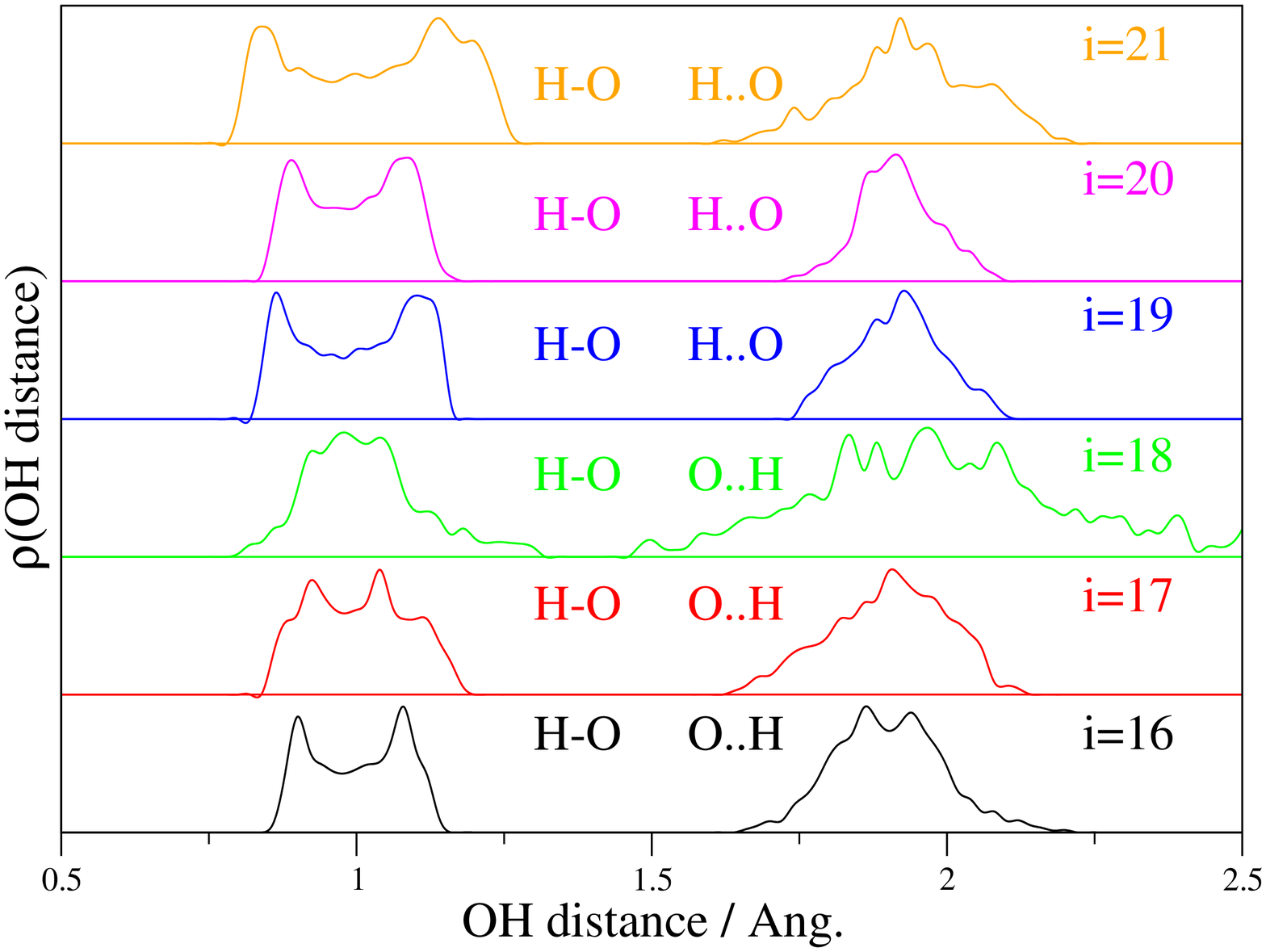}
\par\end{centering}
\caption{\label{fig:Distribution_OH_trimer}Distribution of the OH distance
during a dynamics where the i-th mode is initially excited in order
to enhance its motion. The left side of the panel reports the OH distance
of bound atoms, while in the right side of the panel the distribution
of OH involved in the hydrogen bonds (O...H) is presented. The different
colors refer to different modes.}
\end{figure}

\begin{figure}
\begin{centering}
\includegraphics[scale=0.5]{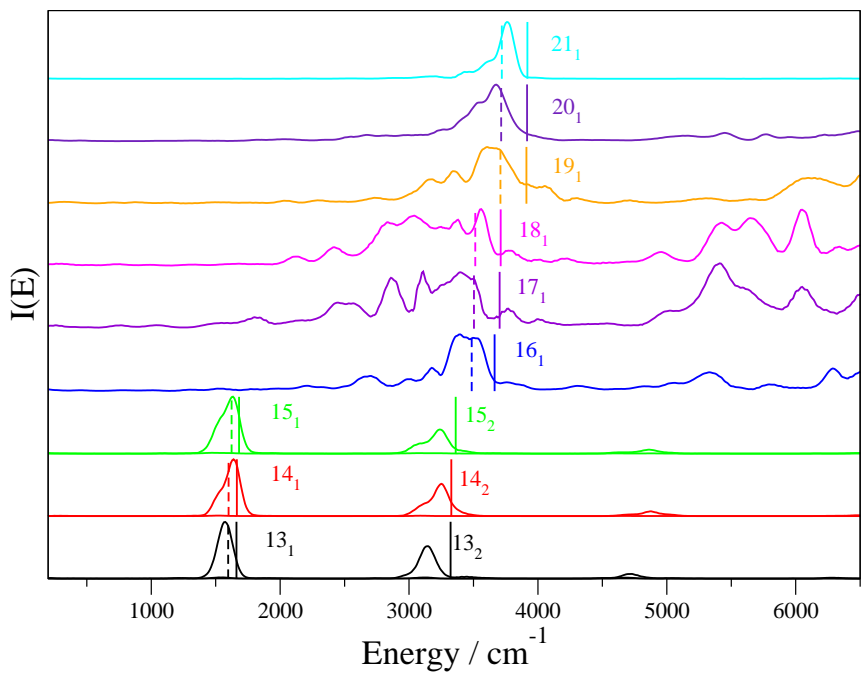}
\par\end{centering}
\caption{\label{fig:Vibrational-spectra-trimer} Vibrational spectra of the
water trimer. The solid lines refer to MC-DC-SCIVR simulations based
on 10 trajectories for each subspace; vertical solid lines indicate
the harmonic estimates, while dashed ones the benchmark MultiMode
values. The bending fundamental and overtone signals were obtained
by tuning the reference state according to Eq. (\ref{eq:MC_reference_state}).}
\end{figure}

The energy range of the vibrational levels is very similar between
the two oligomers so far investigated, with the trimer having the
bending frequencies slightly blue shifted with respect to the dimer,
in agreement with results reported in the literature.\citep{paul_saykally_bendingwaterclusters_1999,schwan_havenith_bendingwaterclusters_2016,wang_bowman_WHBB2_2016}
Larger differences may be found for the OH stretchings which are in
general red shifted with respect to the dimer. Semiclassical results
show a strong red-shift of mode 18 and, more mildly, of modes 16 and
17 with respect to MM values, which is responsible for most of the
MAE values reported in Table (\ref{tab:trimer}). The red shift found
with semiclassical calculations can be explained by looking at Figure
(\ref{fig:Distribution_OH_trimer}), where for modes 16-21 we compare
the distributions of intramolecular (O-H) distances to (O..H) distances
involved in the hydrogen bonds along trajectories with mode-specific
excitation. Modes 19 and 20 seem to be not affected at all by long
range interactions, as expected by their free OH stretching nature.
Mode 21 is also a high-frequency free stretch. It has a broader distribution
of the intramolecular O-H distance which contributes to the appearance
of a tail at shorter distances for the intermolecular O..H distance.
The intramolecular motion for mode 21 is dominant, while the tail
of the distribution is not directional and hydrogen bonding is not
effective. On the opposite, for modes 16, 17, 18 (with particular
emphasis for the latter) the short O..H distances are related to a
strong dynamical hydrogen interaction. Because of it, a bond weakening
is expected resulting into a red-shift of the vibrational frequency.
This is clearly seen upon comparison with MM results. The dynamical
effect that influences modes involved in hydrogen bonds justifies
a large part of the 54 (or 65 for MC-DC-SCIVR) wavenumbers of MAE
with respect to MM values, since if only bending and free OH frequencies
(which are not affected by hydrogen bonding) are compared, then the
MAE reduces to 20 (or 40 for MC-DC-SCIVR) cm$^{-1}$. As anticipated,
single-trajectory MC-DC-SCIVR simulations are not reliable and the
major improvement we have found upon moving to a MC-DC-SCIVR approach
based on multiple minima (and trajectories) concerns the low-energy
red-shifted OH stretches because of the non-local hydrogen interactions.
Such non-local behavior could justify the differences that arise between
our results and MM ones, while another source of discrepancy might
come from the fact we employed a different (even if similar) PES for
the three-body interaction. Furthermore, as already pointed out for
the dimer, these low frequency OH stretches show much more complex
spectral features with respect to modes 13-16 and 19-21 owing to the
interactions with the bending overtones (which are in the same energy
range) or other stretches as depicted by Figure (\ref{fig:Vibrational-spectra-trimer}).
 Figure (\ref{fig:Vibrational-spectra-trimer}) shows the computed
spectra employing the MC-DC-SCIVR approach (solid lines) and reports
MM (dashed vertical lines) and harmonic (solid vertical lines) frequencies.
Bending overtones are very sensitive to the energy of the trajectories
employed in the semiclassical calculations and they cannot be precisely
detected when employing a dynamics energetically tailored on the OH
stretchings, and this adds to the complexity of the resulting spectra.
In particular, mode 18 presents several low-frequency spectral features
due to the overtone bendings and a peak which is blue shifted compared
to the MM frequency and that is due to mode 19.

To point out the importance of a semiclassical approach we have also
computed classical-like spectra obtained from the Fourier transform
of velocity-velocity correlation functions. The trajectory starting
conditions were sampled using the same strategy adopted for semiclassical
simulations with the aim to make the comparison between the different
approaches as straightforward as possibile. In the classical-like
case, 5,000 classical trajectories for each fundamental mode were
enough to get reliable results. We notice that the classical estimates
for the three OH bending frequencies are substantially red-shifted
with respect to MM, LMM, and DC SCIVR ones, while free OH stretches
are found in better agreement. Looking at modes 16-18, the red-shift
is less prominent when the outcomes of classical-like simulations
are compared with DC SCIVR results. Furthermore, as expected, no overtone
features are present in these classical-like spectra.

Finally, we compare in Table (\ref{tab:trimer_exp}) our results to
available experimental data, and find low discrepancies (about 30
cm$^{-1}$ on average) within the typical semiclassical accuracy.
As anticipated, classical-like results are off the mark in the bending
region.
\begin{table}
\centering{}\caption{\label{tab:trimer_exp}Experimental vibrational frequencies available
for the water trimer, in $\text{c\ensuremath{m^{-1}}}$, and calculated
ones. Experimental data are not assigned to a specific vibrational
mode, so for each theoretical approach the closest frequency has been
chosen for comparison. Labels are the same as in previous Tables.
MAE is the mean absolute error referred to experimental data.}
\begin{tabular}{ccccccc}
 & {\scriptsize{}Exp.\citep{huisken_kulcke_watertrimerspectra_1996,schwan_havenith_bendingwaterclusters_2016}} & {\scriptsize{}MM}{\tiny{}\citep{Wang_Bowman_local-monomer_2011}} & {\scriptsize{}LMM}{\tiny{}\citep{Wang_Bowman_local-monomer_2011}} & {\footnotesize{}DC SCIVR$_{\text{10k}}$} & {\footnotesize{}MC-DC SCIVR$_{\text{10trajs,multmin}}$} & $FT(\text{\ensuremath{C_{vv}}})$\tabularnewline
\hline 
 & {\scriptsize{}1608} & {\scriptsize{}1597} & {\scriptsize{}1602} & {\scriptsize{}1584} & {\scriptsize{}1575} & {\scriptsize{}1520}\tabularnewline
\hline 
 & {\scriptsize{}1609} & {\scriptsize{}1600} & {\scriptsize{}1614} & {\scriptsize{}1595} & {\scriptsize{}1637} & {\scriptsize{}1520}\tabularnewline
\hline 
 & {\scriptsize{}1629} & {\scriptsize{}1623} & {\scriptsize{}1615} & {\scriptsize{}1627} & {\scriptsize{}1634} & {\scriptsize{}1516}\tabularnewline
\hline 
 & {\scriptsize{}3533} & {\scriptsize{}3514} & {\scriptsize{}3510} & {\scriptsize{}3640} & {\scriptsize{}3610} & {\scriptsize{}3536}\tabularnewline
\hline 
 & {\scriptsize{}3726} & {\scriptsize{}3720} & {\scriptsize{}3719} & {\scriptsize{}3736} & {\scriptsize{}3760} & {\scriptsize{}3697}\tabularnewline
\hline 
{\scriptsize{}MAE} & - & {\scriptsize{}10} & {\scriptsize{}11} & {\scriptsize{}31} & {\scriptsize{}35} & {\scriptsize{}64}\tabularnewline
\hline 
\end{tabular}
\end{table}

\subsection{Water Hexamer prism (H\protect\textsubscript{2}O)\protect\textsubscript{6}}

In this Section we explore the vibrational features of the water hexamer
prism. It presents 48 degrees of freedom, 18 of which are bendings
and OH stretches. Similarly to what happens moving from the dimer
to the trimer, the magnitude of interactions becomes less intense
going from the trimer to the hexamer. Indeed, with the same Hessian
threshold value adopted for the trimer, all the modes of the hexamer
have been treated independently, with the exception of a couple of
modes (number 35 and 45) which have been still enrolled into a bi-dimensional
subspace.

We computed the hexamer DC-SCIVR spectra with 5,000 trajectories per
subspace, and, similarly to the case of the dimer and trimer, we also
checked the reliability of a MC-DC-SCIVR approach based on just a
few trajectories. However, from our damped dynamics simulations it
was soon evident that too many local minima had to be taken into consideration.
Running a trajectory from each minimum would have not provided a real
computational advantage over DC-SCIVR, so we introduced a different
approach to select the most relevant minima for our calculations. 

We adopted a strategy inspired by Habershon's recent work on correlation
distributions.\citep{Habershon_correlation_2016} As anticipated,
the many-body PES of the hexamer is characterized by several local
minima, very close in energy to each other. To assess the ``vicinity''
of each minimum to the global one, we chose a connection criterion
based on structural considerations. Specifically, we introduced a
correlation parameter computed as a sum of molecular distances. In
fact, by defining a proper set of distances $\text{\ensuremath{\left\{  d_{i}\right\}  _{i=1,\ldots,N_{distances}}}}$
between the atoms, the correlation parameter $\sigma^{2}$ can be
calculated as 
\begin{equation}
\sigma^{2}=\sum_{i=1}^{N_{distances}}(d-d_{i}^{ref})^{2},\label{eq:correlation_formula}
\end{equation}
where $\text{\ensuremath{d_{i}^{ref}}}$ is the i-th distance calculated
at the global minimum geometry. In the set of distances we included
the length of OH bonds for each of the six monomers plus the O-O distances
of adjacent oxygen atoms. By looking at Eq. (\ref{eq:correlation_formula})
it is expected that minima very correlated to the global one have
$\sigma^{2}\rightarrow0$, while higher values identify less correlated
wells. 

\begin{figure}
\begin{centering}
\includegraphics[scale=0.4]{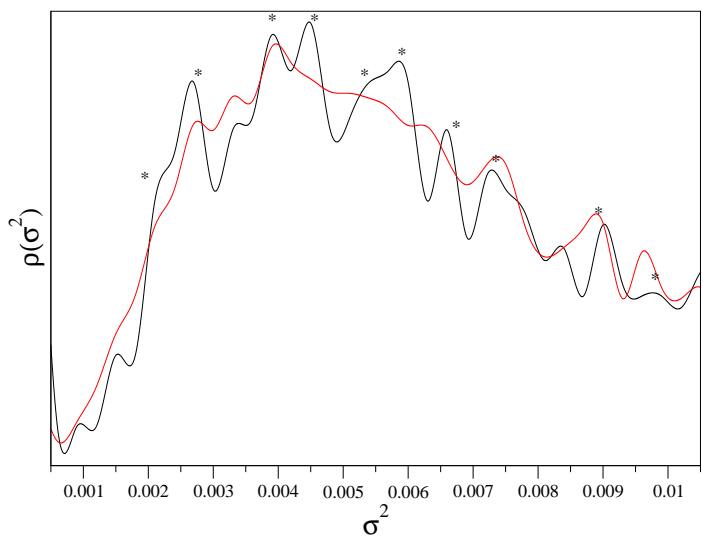}
\par\end{centering}
\centering{}\caption{\label{fig:Correlation_hexamer}Correlation distribution between the
global minimum and the local minima found by exploring the hexamer
many-body PES. The plots report the results for 1,000 (black) and
10,000 (red) damped-dynamics trajectories. Data have been interpolated
by means of a cubic spline. The black asterisks represent the correlation
peaks corresponding to the minima employed in the MC-DC-SCIVR calculations.}
\end{figure}

We took into consideration $\sigma^{2}$ values up to 0.011 $\textrm{Å\ensuremath{^{2}}}$,
which is enough to cover more than 80\% of the located local minima.
Those which are further away from the global minimum (in $\sigma^{2}$
terms) are expected to contribute less significantly to the spectral
features, and are neglected. Figure (\ref{fig:Correlation_hexamer})
shows the correlation distribution as a function of $\sigma^{2}$
for different sets of damped-dynamics trajectories employed. As in
previous cases, the initial conditions of the damped trajectories
are sampled by means of a Husimi distribution around the PES global
minimum, and a damping factor equal to 0.99 is adopted. Peaks in the
distribution, sampled along the $\sigma^{2}$ range studied, are used
to select the most relevant minima for the MC-SCIVR calculations. 

\begin{table}
\centering{}\caption{\label{tab:hexamer}Vibrational frequencies of the water hexamer prism,
in $\text{c\ensuremath{m^{-1}}}.$ The first column refers to the
mode-excitation label; the second column is the harmonic estimate;
the third column reports the Local Monomer Model results; from the
fourth column on, the semiclassical values are listed. The MC-DC-SCIVR
simulation based on 11 trajectories has been performed upon selection
of 10 local minima from the correlation distribution. MAE values are
referred to LMM ones.}
{\tiny{}}%
\begin{tabular}{cccccc}
{\footnotesize{}Index} & {\footnotesize{}HO} & {\footnotesize{}LMM\citep{Wang_Bowman_local-monomer_2011}} & {\footnotesize{}DC SCIVR$_{\text{5k}}$} & {\footnotesize{}MC-DC SCIVR\textsubscript{{\tiny{}11trajs,multmin}}} & {\footnotesize{}MC-DC SCIVR\textsubscript{1traj}}\tabularnewline
\hline 
{\footnotesize{}$31_{1}$} & {\footnotesize{}1661} & {\footnotesize{}1606} & {\footnotesize{}1617} & {\footnotesize{}1602} & {\footnotesize{}1606}\tabularnewline
\hline 
{\footnotesize{}$32_{1}$} & {\footnotesize{}1672} & {\footnotesize{}1612} & {\footnotesize{}1623} & {\footnotesize{}1620} & {\footnotesize{}1592}\tabularnewline
\hline 
{\footnotesize{}$33_{1}$} & {\footnotesize{}1676} & {\footnotesize{}1620} & {\footnotesize{}1622} & {\footnotesize{}1622} & {\footnotesize{}1588}\tabularnewline
\hline 
{\footnotesize{}$34_{1}$} & {\footnotesize{}1701} & {\footnotesize{}1633} & {\footnotesize{}1664} & {\footnotesize{}1636} & {\footnotesize{}1682}\tabularnewline
\hline 
{\footnotesize{}$35_{1}$} & {\footnotesize{}1715} & {\footnotesize{}1654} & {\footnotesize{}1661} & {\footnotesize{}1640} & {\footnotesize{}1684}\tabularnewline
\hline 
{\footnotesize{}$36_{1}$} & {\footnotesize{}1739} & {\footnotesize{}1677} & {\footnotesize{}1715} & {\footnotesize{}1712} & {\footnotesize{}1722}\tabularnewline
\hline 
{\footnotesize{}$37_{1}$} & {\footnotesize{}3377} & {\footnotesize{}3092} & {\footnotesize{}2925} & {\footnotesize{}2956} & {\footnotesize{}3011}\tabularnewline
\hline 
{\footnotesize{}$38_{1}$} & {\footnotesize{}3494} & {\footnotesize{}3256} & {\footnotesize{}3052} & {\footnotesize{}3060} & {\footnotesize{}3012}\tabularnewline
\hline 
{\footnotesize{}$39_{1}$} & {\footnotesize{}3619} & {\footnotesize{}3372} & {\footnotesize{}3182} & {\footnotesize{}3168} & {\footnotesize{}2940}\tabularnewline
\hline 
{\footnotesize{}$40_{1}$} & {\footnotesize{}3638} & {\footnotesize{}3442} & {\footnotesize{}3516} & {\footnotesize{}3395} & {\footnotesize{}3198}\tabularnewline
\hline 
{\footnotesize{}$41_{1}$} & {\footnotesize{}3714} & {\footnotesize{}3482} & {\footnotesize{}3573} & {\footnotesize{}3556} & {\footnotesize{}3200}\tabularnewline
\hline 
{\footnotesize{}$42_{1}$} & {\footnotesize{}3735} & {\footnotesize{}3521} & {\footnotesize{}3640} & {\footnotesize{}3616} & {\footnotesize{}3500}\tabularnewline
\hline 
{\footnotesize{}$43_{1}$} & {\footnotesize{}3792} & {\footnotesize{}3579} & {\footnotesize{}3592} & {\footnotesize{}3606} & {\footnotesize{}3680}\tabularnewline
\hline 
{\footnotesize{}$44_{1}$} & {\footnotesize{}3809} & {\footnotesize{}3588} & {\footnotesize{}3580} & {\footnotesize{}3574} & {\footnotesize{}3608}\tabularnewline
\hline 
{\footnotesize{}$45_{1}$} & {\footnotesize{}3827} & {\footnotesize{}3630} & {\footnotesize{}3678} & {\footnotesize{}3650} & {\footnotesize{}3602}\tabularnewline
\hline 
{\footnotesize{}$46_{1}$} & {\footnotesize{}3915} & {\footnotesize{}3697} & {\footnotesize{}3771} & {\footnotesize{}3610} & {\footnotesize{}3578}\tabularnewline
\hline 
{\footnotesize{}$47_{1}$} & {\footnotesize{}3923} & {\footnotesize{}3706} & {\footnotesize{}3698} & {\footnotesize{}3750} & {\footnotesize{}3768}\tabularnewline
\hline 
{\footnotesize{}$48_{1}$} & {\footnotesize{}3925} & {\footnotesize{}3728} & {\footnotesize{}3677} & {\footnotesize{}3712} & {\footnotesize{}3700}\tabularnewline
\hline 
{\footnotesize{}MAE} & {\footnotesize{}169} & {\footnotesize{}-} & {\footnotesize{}64} & {\footnotesize{}57} & {\footnotesize{}102}\tabularnewline
\end{tabular}
\end{table}

\begin{figure}
\begin{centering}
\includegraphics[scale=0.5]{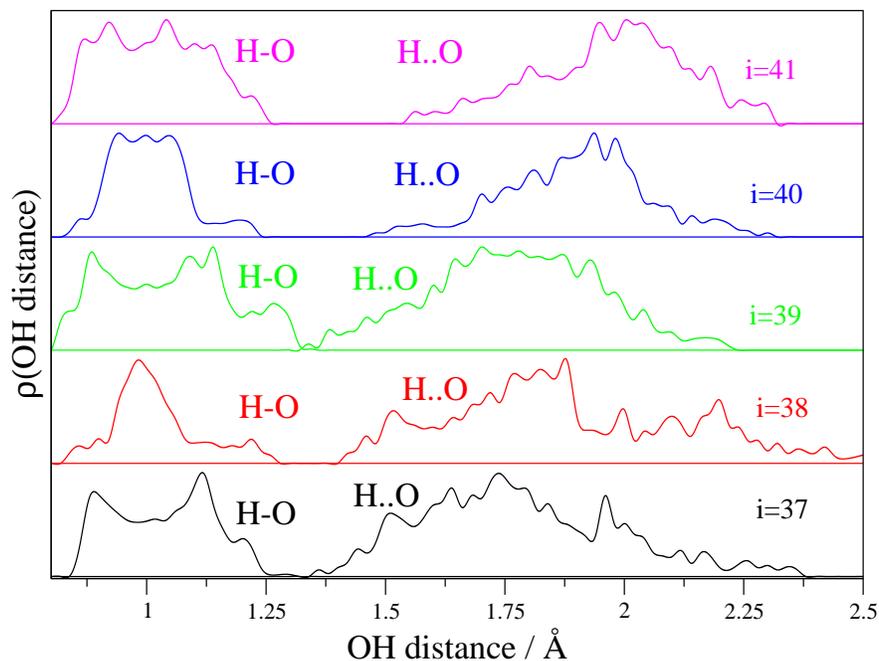}
\par\end{centering}
\caption{\label{fig:Distribution_OH_hexamer}Distribution of the OH distance
during a dynamics where the i-th mode is initially excited in order
to enhance its motion. The left side of the panel reports the OH distance
of bound atoms, while in the right side of the panel the distribution
of OH involved in the hydrogen bonds (O...H) is presented. The different
colors refer to different modes.}
\end{figure}

Table (\ref{tab:hexamer}) shows our results compared to the Local
Monomer Model ones.\citep{Wang_Bowman_local-monomer_2011} Vibrations
are predicted by LMM at lower frequencies than the typical OH stretching
region, approximately in the range between 3,100 and 3,300 cm\textsuperscript{-1}.
Some semiclassical results were found in the 2,900-3,100 cm\textsuperscript{-1}
region of the spectrum. Therefore our dynamics-based results for the
lower-frequency stretches are red shifted with respect to the time-independent
ones. We ascribe the main reason for this discrepancy to the different
impact of hydrogen interactions, which weaken such OH bonds. In a
dynamical approach, the effect of hydrogen bonding is enhanced, as
already evinced for the trimer. A further evidence of this is reported
in Figure (\ref{fig:Distribution_OH_hexamer}) where we compare the
distributions of intramolecular (O-H) distances to (O..H) distances
involved in the hydrogen bonds for modes 37-41 along trajectories
with specific mode excitation. Modes 37-39, which are semiclassically
the most red-shifted ones, present a more prominent tail at short
O..H distances with respect to modes 40-41, which are not red shifted.
These features point to a stronger OH..O hydrogen interaction for
modes 37-39 than for the other stretches. Consequently, the corresponding
OH bonds are weakened and their frequencies red shifted. MAE values
relative to the LMM ones are around 60 cm\textsuperscript{-1}for
DC-SCIVR and MC-DC-SCIVR based on 11 trajectories, while the MAE is
substantially higher (\textasciitilde{}100 cm\textsuperscript{-1})
when a single trajectory is employed. These values are much lower
if only modes not involved in hydrogen bonding, i.e. modes 31-36 and
43-48, are considered. In fact the MAE decreases to 25 cm\textsuperscript{-1}for
DC SCIVR, and to 22 cm\textsuperscript{-1}for MC-DC SCIVR with 11
trajectories. 

\begin{figure}
\begin{centering}
\includegraphics[scale=0.5]{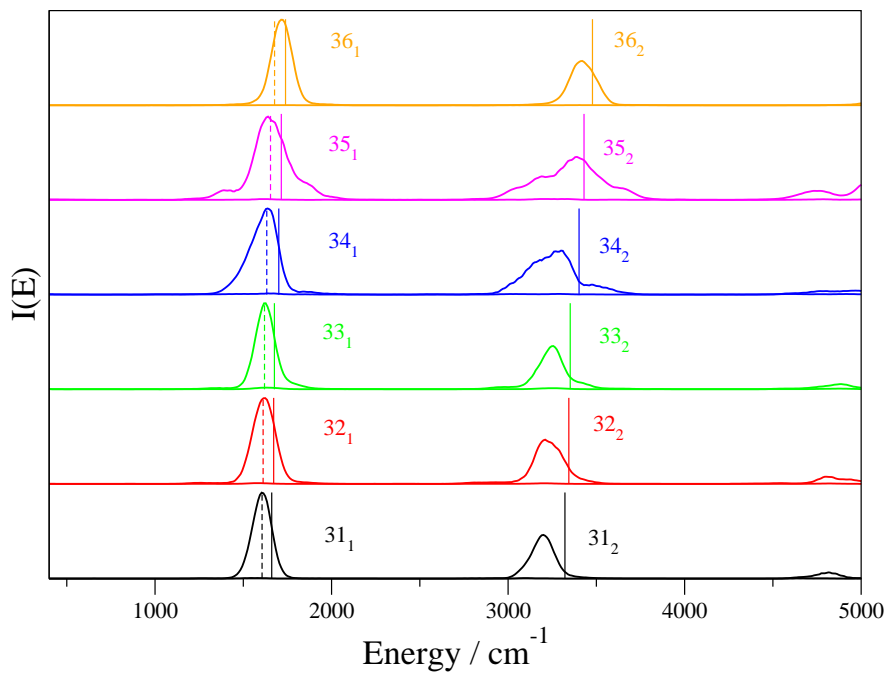}
\par\end{centering}
\caption{\label{fig:Vibrational-spectra-hexamer}Vibrational spectra of the
water hexamer prisms in the bending region. The solid lines refer
to MC-DC-SCIVR simulations based on 11 trajectories for each subspace;
vertical solid lines indicate the harmonic estimates, while dashed
ones the Local Monomer values. The bending fundamental and overtone
signals were obtained by tuning the reference state according to Eq.
(\ref{eq:MC_reference_state}).}
\end{figure}

\begin{figure}
\begin{centering}
\includegraphics[scale=0.5]{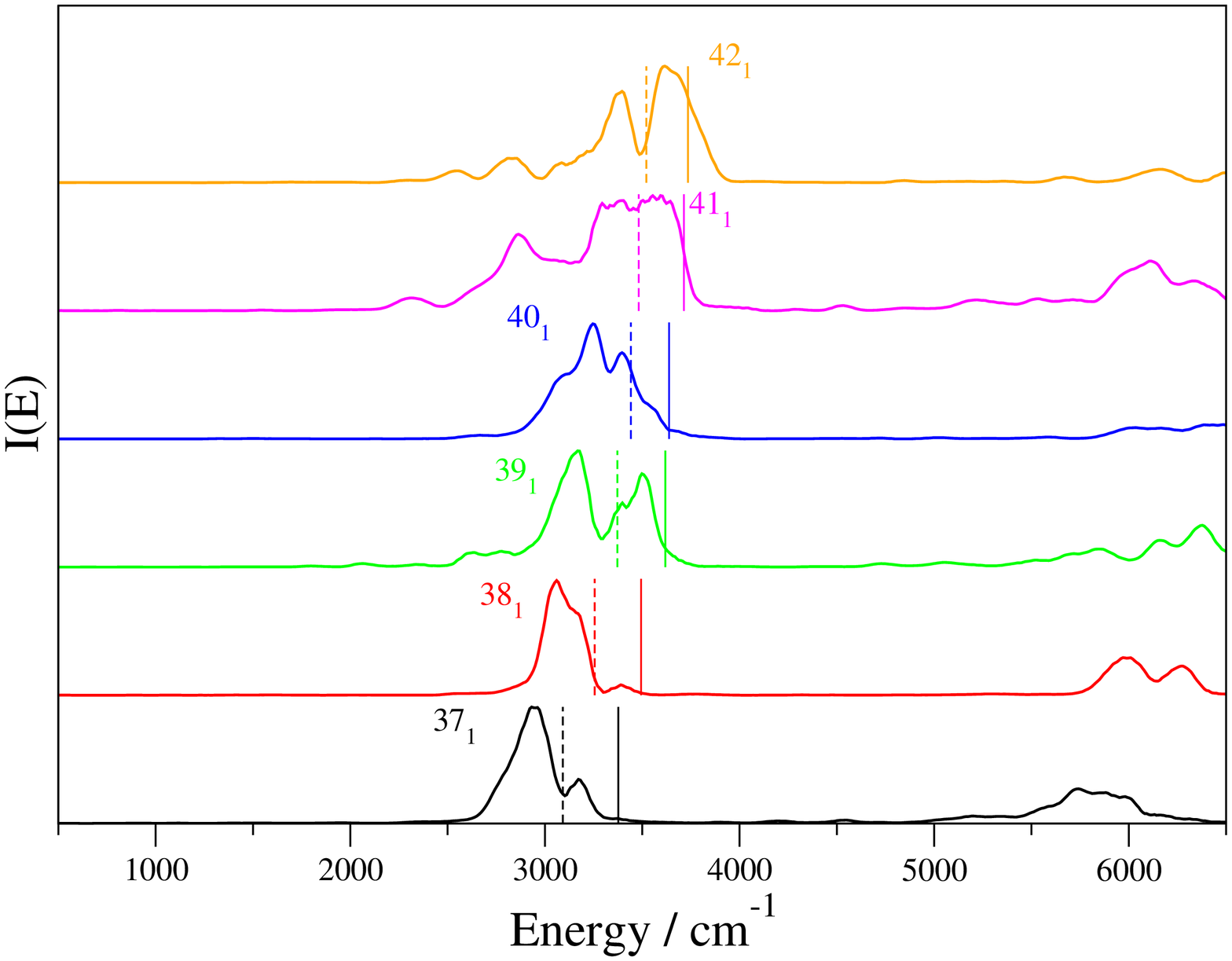}
\par\end{centering}
\caption{\label{fig:Vibrational-spectra-hexamer2}Vibrational spectra of the
water hexamer prism in the OH stretching region. The solid lines refer
to MC-DC-SCIVR simulations based on 11 trajectories for each subspace;
vertical solid lines indicate the harmonic estimates, while dashed
ones the Local Monomer values.}
\end{figure}

\begin{figure}
\begin{centering}
\includegraphics[scale=0.5]{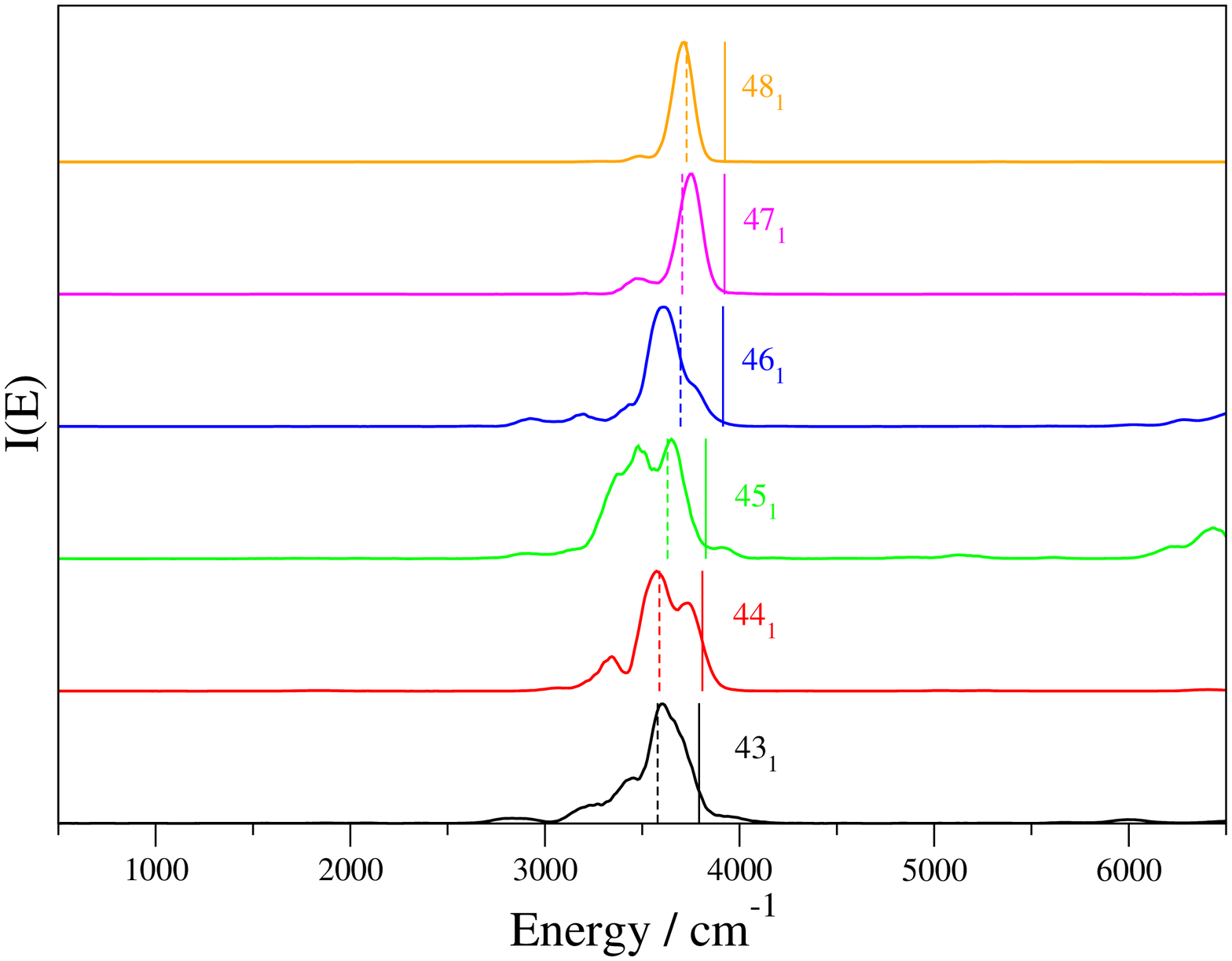}
\par\end{centering}
\caption{\label{fig:vibrational-spectra-hexamer3}Vibrational spectra of the
water hexamer prism in the free OH stretching region. The solid lines
refer to MC-DC-SCIVR simulations based on 11 trajectories for each
subspace; vertical solid lines indicate the harmonic estimates, while
dashed ones the Local Monomer values. }
\end{figure}

Figures (\ref{fig:Vibrational-spectra-hexamer}), (\ref{fig:Vibrational-spectra-hexamer2}),
and (\ref{fig:vibrational-spectra-hexamer3}) show the spectra computed
for the hexamer employing the MC-DC-SCIVR approach with 11 trajectories
(solid line) together with the LMM (dashed vertical lines) and harmonic
(solid vertical lines) frequency estimates. Specifically, in Figure
(\ref{fig:Vibrational-spectra-hexamer}) the six bendings and their
overtones are reported, while Figures (\ref{fig:Vibrational-spectra-hexamer2})
and (\ref{fig:vibrational-spectra-hexamer3}) are dedicated to the
low-frequency and free OH stretches respectively. If, for bendings,
spectral features are well resolved, peaks associated to the stretches
are instead broader and have a more complex shape due to the intermode
couplings involving both stretches and overtones of bendings. This
was also observed in the trimer and it is evident in the hexamer too.
In particular, power spectra of modes 39 and 40 present a double peak
feature due to the coupling between these two modes. A similar instance
occurs for spectra of modes 41 and 42 which, in addition, show a shoulder
at the frequency of mode 37. The effects of coupling fade away when
moving to the free OH stretches, a characteristic which has been already
found in the trimer.


In summary, the semiclassical information about the hexamer energy
levels based on thousands of trajectories can be basically regained
by means of a MC-DC-SCIVR treatment that employs just 11 selected
trajectories. A single-trajectory approach is instead not enough to
recover the correct spectral features due to the strong influence
of local minima similar in both energy and connectivity to the global
one. Therefore, our MC-DC-SCIVR approach is very promising for dealing
with higher-dimensional clusters for which a DC-SCIVR calculation
is out of reach as in the case of the water decamer $\left(\text{H}_{2}\text{O}\right)_{10}$.

\subsection{Water Decamer (H\protect\textsubscript{2}O)\protect\textsubscript{10}}

Our last application concerns the water decamer which has 10 bendings
and 20 OH stretches, and a total of 84 vibrational degrees of freedom.
Due to the computational overhead of the simulations, for the decamer
we only performed MC-DC-SCIVR calculations based on multiple trajectories
starting from a set of minima located by means of the damped-dynamics
approach. We identified several minima on the surface and calculated
a correlation distribution dependent on intramolecular OH distances
and OO distances of adjacent monomers according to Eq. (\ref{eq:correlation_formula}),
from which we extracted the 10 most relevant local minima. By employing
the same Hessian threshold value adopted for the trimer and the hexamer,
all 30 degrees of freedom have been treated as independent ones. This
is in agreement with the trend of weakening interactions between vibrational
modes associated with an increase in the dimensionality of the system.
For each subspace we performed our MC-DC-SCIVR calculations based
on 11 trajectories initiated from the global minimum and the 10 chosen
local minima. Results are reported in Table (\ref{tab:decamer}) which
shows a comparison of the decamer semiclassical fundamental frequencies
with the corresponding values calculated with the Local Monomer Model.\citep{Wang_Bowman_local-monomer_2011}

\begin{table}
\centering{}\caption{\label{tab:decamer}Fundamental frequencies of vibration for the water
decamer ($cm^{-1}$). HO is the label for the harmonic frequencies;
LMM are the Local Monomer Model results; MC-DC-SCIVR refers to our
semiclassical estimates. }
\begin{tabular}{ccccccc}
{\footnotesize{}HO} & {\footnotesize{}LMM\citep{Wang_Bowman_local-monomer_2011}} & {\footnotesize{}MC-DC-SCIVR\textsubscript{{\tiny{}11 trajs, multmin}}} &  & {\footnotesize{}HO} & {\footnotesize{}LMM\citep{Wang_Bowman_local-monomer_2011}} & {\footnotesize{}MC-DC-SCIVR\textsubscript{{\tiny{}11 trajs, multmin}}}\tabularnewline
\hline 
{\footnotesize{}1670} & {\footnotesize{}1600} & {\footnotesize{}1590} &  & {\footnotesize{}3571} & {\footnotesize{}3382} & {\footnotesize{}3337}\tabularnewline
\hline 
{\footnotesize{}1675} & {\footnotesize{}1602} & {\footnotesize{}1624} &  & {\footnotesize{}3659} & {\footnotesize{}3417} & {\footnotesize{}3379}\tabularnewline
\hline 
{\footnotesize{}1678} & {\footnotesize{}1608} & {\footnotesize{}1624} &  & {\footnotesize{}3666} & {\footnotesize{}3419} & {\footnotesize{}3400}\tabularnewline
\hline 
{\footnotesize{}1686} & {\footnotesize{}1609} & {\footnotesize{}1628} &  & {\footnotesize{}3676} & {\footnotesize{}3420} & {\footnotesize{}3406}\tabularnewline
\hline 
{\footnotesize{}1692} & {\footnotesize{}1617} & {\footnotesize{}1660} &  & {\footnotesize{}3682} & {\footnotesize{}3429} & {\footnotesize{}3448}\tabularnewline
\hline 
{\footnotesize{}1712} & {\footnotesize{}1647} & {\footnotesize{}1663} &  & {\footnotesize{}3727} & {\footnotesize{}3518} & {\footnotesize{}3492}\tabularnewline
\hline 
{\footnotesize{}1713} & {\footnotesize{}1664} & {\footnotesize{}1674} &  & {\footnotesize{}3741} & {\footnotesize{}3525} & {\footnotesize{}3496}\tabularnewline
\hline 
{\footnotesize{}1720} & {\footnotesize{}1665} & {\footnotesize{}1690} &  & {\footnotesize{}3756} & {\footnotesize{}3534} & {\footnotesize{}3522}\tabularnewline
\hline 
{\footnotesize{}1738} & {\footnotesize{}1669} & {\footnotesize{}1708} &  & {\footnotesize{}3774} & {\footnotesize{}3566} & {\footnotesize{}3532}\tabularnewline
\hline 
{\footnotesize{}1748} & {\footnotesize{}1691} & {\footnotesize{}1714} &  & {\footnotesize{}3781} & {\footnotesize{}3568} & {\footnotesize{}3565}\tabularnewline
\hline 
{\footnotesize{}3335} & {\footnotesize{}3013} & {\footnotesize{}2936} &  & {\footnotesize{}3914} & {\footnotesize{}3706} & {\footnotesize{}3640}\tabularnewline
\hline 
{\footnotesize{}3352} & {\footnotesize{}3036} & {\footnotesize{}3006} &  & {\footnotesize{}3920} & {\footnotesize{}3734} & {\footnotesize{}3668}\tabularnewline
\hline 
{\footnotesize{}3383} & {\footnotesize{}3046} & {\footnotesize{}3022} &  & {\footnotesize{}3924} & {\footnotesize{}3736} & {\footnotesize{}3672}\tabularnewline
\hline 
{\footnotesize{}3387} & {\footnotesize{}3050} & {\footnotesize{}3052} &  & {\footnotesize{}3925} & {\footnotesize{}3741} & {\footnotesize{}3680}\tabularnewline
\hline 
{\footnotesize{}3554} & {\footnotesize{}3286} & {\footnotesize{}3121} &  & {\footnotesize{}3926} & {\footnotesize{}3744} & {\footnotesize{}3800}\tabularnewline
\hline 
\end{tabular}
\end{table}
In this case we observe that our results for the bendings are generally
in good agreement with LMM ones with a MAE equal to 36 wavenumbers.
Both MC-DC-SCIVR and LMM predict more red-shifted stretches than in
the case of the smaller clusters, but they are in closer agreement
with respect to the trimer or the hexamer.

\section{Summary and Conclusions\label{sec:Conclusions}}

In this paper we have presented a semiclassical investigation of the
vibrational features of some water clusters ranging from the dimer
to the decamer by means of our recently established Divide-and-Conquer
semiclassical approach. Semiclassical simulations employ several thousand
classical trajectories to reach convergence of results, but a computationally-cheaper
MC-DC-SCIVR approach based on few, selected trajectories was demonstrated
to provide quite acceptable results. The caveat here is that, differently
from other molecular systems studied in the past, a single-minimum/single-trajectory
semiclassical calculation is usually not accurate. Therefore, we have
explored the potential energy surface looking for local minima and
presented a way to select them according to their ``resemblance''
to the global minimum and their expected contribution to the calculations.
The application of semiclassical methods to water clusters demonstrates
that these techniques can be employed also for large $\left(\text{H}_{2}\text{O}\right)_{n}$
ones as well as for rather floppy systems, and not only for quite
rigid ones. The divide-and-conquer method is able to simplify the
full-dimensional problem recovering part of the interactions between
the low-dimensional subspaces thanks to the maintained full-dimensional
nature of the trajectories on which the subspace calculations are
based. Spectral features though are very sensitive to intermode couplings
and multiple peak structures are often present especially in the case
of low frequency stretches. Furthermore, vibrational angular momentum
due to the floppy nature of the system contributes to increase the
width of peaks, which is substantially larger than what is commonly
found in semiclassical calculations of single molecules. 

Results show that the outcomes of experiments and previous theoretical
studies are regained with quantitative agreement for bendings and
free OH stretches, while frequencies of OH stretches influenced by
hydrogen-bond interactions are red shifted with respect to the estimates
provided by other theoretical approaches. This can be clearly seen
in the assignment of the trimer experimental frequency at 3533 cm\textsuperscript{-1}.
We assign it semiclassically to mode 19, while VCI calculations yield
a closer estimate for the frequency of mode 18, and classical-like
simulations point to mode 16. The difference between semiclassical
and classical-like estimates is evident and confirms the need to undertake
a semiclassical approach able to regain quantum effects. The presence
of a set of semiclassical frequencies around 3,000 cm\textsuperscript{-1}
for the hexamer and the decamer is consistent with previous studies
even if the red shift is more accentuated in our simulations. This
is due to dynamical effects (confirmed by the short-distance tails
of the O..H distance distributions for modes involved in hydrogen
bonds) and to the multi-reference nature of the semiclassical approach.
Compared to the isolated water molecule, bending frequencies are more
and more blue shifted and low-frequency stretches more and more red-shifted
as the cluster size increases. Agreement between semiclassical and
VCI calculations for modes in the red-shift region is better for the
decamer than for smaller clusters. 

\section*{SUPPLEMENTARY MATERIAL}

See Supplementary Material for the new 3-body water-water-water PES
employed in this work.

\section*{Acknowledgments}

Professor Joel M. Bowman is warmly acknowledged for providing his
WHBB Potential Energy Surface and for very useful discussions and
suggestions. R.C. thanks Prof. Bowman and Dr. Yimin Wang for providing
the WHBB 3-body ab initio energy database employed to construct the
fast water-water-water interaction potential adopted in this work.
Authors acknowledge support from the European Research Council (ERC)
under the European Union\textquoteright s Horizon 2020 research and
innovation programme (grant agreement No {[}647107{]} \textendash{}
SEMICOMPLEX \textendash{} ERC-2014-CoG). M.C. acknowledges also the
CINECA and the Regione Lombardia award under the LISA initiative (grant
LI05p\_GREENTI\_0) for providing high perfomance computational resources
at CINECA (Italian Supercomputing Center). All authors thank Università
degli Studi di Milano for further computational time at CINECA.

\newpage

\bibliographystyle{aipnum4-1}
\bibliography{SEMICOMPLEX}

\end{document}